\documentclass[12pt,letterpaper]{article}

\usepackage{amscd,amsmath,amssymb,amsfonts,xspace,mathrsfs}
\usepackage{hyperref} 
\usepackage[multiple]{footmisc}
\usepackage{url}

\numberwithin{equation}{section}

\def\varpi{t}

\def\sign{{\rm sign}}

\def\Im{\,{\rm Im}\,}

\def\({\left(}
\def\){\right)}
\def\[{\left[}
\def\]{\right]}
\def\<{\left\langle}
\def\>{\right\rangle}

\newcommand{\half}{\frac{1}{2}}

\newcommand{\I}{\mathrm{i}}

\newcommand{\cM}{\mathcal{M}}

\newcommand{\non}{\nonumber}
\DeclareSymbolFont{AMSa}{U}{msa}{m}{n}
\DeclareSymbolFont{AMSb}{U}{msb}{m}{n}
\DeclareMathSymbol{\fieldR}{\mathalpha}{AMSb}{"52}

\newcommand{\eps}{\epsilon}

\newcommand{\IR}{\mathbb{R}}
\newcommand{\IC}{\mathbb{C}}
\newcommand{\IZ}{\mathbb{Z}}
\newcommand{\IH}{\mathbb{H}}

\newcommand{\IP}{\mathbb{P}}
\newcommand{\Zint}{\mathbb{Z}}

\newcommand{\sgn}{\mbox{sgn}}

\def\bea{\begin{eqnarray}}
\def\eea{\end{eqnarray}}
\def\be{\begin{equation}}
\def\ee{\end{equation}}
\def\ba{\begin{align}}
\def\ea{\end{align}}
\def\bse{\begin{subequations}}
\def\ese{\end{subequations}}

\newcommand{\bfk}{{\boldsymbol k}}

\newcommand{\bfm}{{\boldsymbol m}}
\newcommand{\bfp}{{\boldsymbol p}}

\newcommand{\bft}{{\boldsymbol t}}

\def\ba{\bar a}

\def\cij#1{c}
\def\ci#1{c}

\def\gamD#1{\tilde\gamma}

\def\cl0{\tilde c_0}

\newcommand{\noi}{\noindent}

\def\bfu{\boldsymbol{u}}
\def\bfv{\boldsymbol{v}}

\begin{document}

\begin{titlepage}

\begin{center}

\hfill TCD-17-17
{}
\vskip 2cm {\LARGE \bf Vafa-Witten theory and iterated integrals
  \newline of
modular forms} \vskip
1.25 cm {Jan Manschot$^{1,2}$}\\
\vskip 0.5cm  
$^1$ {\it School of Mathematics, Trinity College, Dublin 2, Ireland}\\
\vskip .3cm
$^2$ {\it Hamilton Mathematical Institute, Trinity College, Dublin 2, Ireland}\\

\end{center}

\vskip 2 cm

\begin{abstract}
\baselineskip=18pt 
Vafa-Witten (VW) theory is a topologically twisted version of
$\mathcal{N}=4$ supersymmetric Yang-Mills theory. $S$-duality suggests
that the partition function of VW theory with gauge group $SU(N)$ transforms as
a modular form under duality transformations. Interestingly,  Vafa and
Witten demonstrated the presence of a modular anomaly, when
the theory has gauge group $SU(2)$ and is
considered on the complex projective plane $\mathbb{P}^2$. This modular
anomaly could be expressed as an integral of a modular
form, and also be traded for a holomorphic anomaly. We demonstrate that the
modular anomaly for gauge group $SU(3)$ involves an iterated integral
of modular forms. Moreover, the modular anomaly for $SU(3)$ can be traded for a
holomorphic anomaly, which is shown to factor into a product of the
partition functions for lower rank gauge groups. The $SU(3)$ partition function
is mathematically an example of a mock modular form of depth two. 
\noi  
\end{abstract}

\end{titlepage} 
 
\pagestyle{plain} 
\baselineskip=19pt

\tableofcontents

\section{Introduction}
Topological field theory has been important for establishing symmetries
and dualities in field theory and string theory beyond the semi-classical level
\cite{Witten:1988ze, Bershadsky:1993cx, Vafa:1994tf}. We 
consider in this article the Vafa-Witten twist of $\mathcal{N}=4$ supersymmetric
Yang-Mills (YM) theory, or Vafa-Witten (VW) theory for short. A
generalization of electric-magnetic duality to Yang-Mills theory, $S$-duality,
acts naturally on VW theory \cite{Vafa:1994tf}. This duality, proposed by Montonen and
Olive \cite{Montonen:1977sn}, states that YM theory with gauge group
$G$ and complexified coupling constant
\be
\label{tau}
\tau=\frac{\theta}{2\pi}+\frac{4\pi i}{g^2},
\ee
 has a dual description as YM theory, whose gauge group is the
 Langlands dual group of $G$, $^LG$, and with inverse coupling constant $-1/\tau$. 
Together with the periodicity of the $\theta$-angle, this generates
the $SL(2,\IZ)$ $S$-duality group. Since the unitary groups $U(N)$ are self-dual under Langlands duality,
$S$-duality suggests that the partition function of VW theory transforms as a modular form,  
\be 
\label{ZMN}
Z_{N}\!\left(\frac{a\tau+b}{c\tau+d}\right)\sim Z_{N}(\tau), \quad \qquad
\left(\begin{array}{cc} a & b \\ c & d \end{array} \right)\in SL(2,\IZ).
\ee
where $\sim$ indicates a possible pre-factor, which may depend
polynomially on $\tau$. We will
discuss in some more detail later the richer structure of $SU(N)$
theories under the action of $SL(2,\IZ)$.
   
The suggested transformation (\ref{ZMN}) is hard to verify for a general
four-manifold and gauge group. For a few four-manifolds, such as $K3$ \cite{Vafa:1994tf} and
the rational elliptic surface $\tfrac{1}{2}K3$ \cite{Minahan:1998vr, Yoshioka:1998ti}, modularity of the partition
functions could be verified using string
dualities for $U(N)$ with $N$ arbitrary. The partition functions could in these cases be expressed in terms of classical
modular forms and theta series, and exhibited a rather rich
structure. Depending on the four-manifold, connections to Hecke
operators and quasi-modular forms were made \cite{Minahan:1998vr}.  
 For some other four-manifolds such as the complex projective plane $\mathbb{P}^2$, verification was
limited to gauge group $SU(2)$ and $SO(3)$ \cite{Vafa:1994tf,
  Gottsche:1999ix, Alim:2010cf, Manschot:2011dj}, which motivated the
present work on gauge groups $SU(3)$ and $U(3)$. 

Interestingly, for the complex projective plane
$\mathbb{P}^2$ and other rational surfaces with $b_2^+=1$, the partition function for
gauge group $SU(2)$ is expressed in terms of
functions whose modular properties are more subtle than those of the
classical modular forms. In particular, Vafa and Witten \cite{Vafa:1994tf}
expressed the VW partition function for $\IP^2$ in terms of Zagier's generating function of
class numbers \cite{Zagier:1975}, whose modular transformations deviate from the form in Equation
(\ref{ZMN}). More precisely, its transformations include a shift by an
 integral of a modular form, such as
\be
\label{intTheta0}
\int_{\frac{d}{c}}^{i\infty} \frac{\Theta_0(u)}{(-i(\tau+u))^\frac{3}{2}}du,
\ee
where $\Theta_0$ is a weight $\frac{1}{2}$ theta series defined in Equation (\ref{Theta_alpha}).
The shift by such a period integral is one of the defining properties of
functions known as mock modular forms \cite{Zwegers-thesis,
  MR2605321}. The modular anomaly of such functions can be traded for a
holomorphic anomaly, by including a non-holomorphic period integral to
the partition function. Such an integral is as in Equation
(\ref{intTheta0}), but with $\frac{d}{c}$ replaced by $-\bar
\tau$. Besides their appearance in VW partition functions, mock 
modular forms have become an important element of the study of modular partition functions, such as in conformal field theory \cite{Eguchi:1988af, Semikhatov:2003uc,
  Troost:2010ud}, Donaldson-Witten theory \cite{Moore:1997pc,
  Malmendier:2008db, Korpas:2017qdo}, AdS$_3$ gravity
\cite{Manschot:2007ha}, black holes \cite{Manschot:2009ia,
  Dabholkar:2012nd, Alexandrov:2016tnf} and the
moonshine phenomenon \cite{Cheng:2011ay}.
  
We will show in this paper that for VW theory with gauge group
$SU(3)$, the modular transformations of the partition functions include a shift
by iterated (double) integrals of theta series. One instance of these integrals is:
\be
\label{itintegral}
\int_{\frac{d}{c}}^{i\infty} \int_{u_2}^{i\infty}
\frac{\Theta_0(u_1)\,\Theta_{0}(3u_2)}{\sqrt{-(u_1+\tau)^3
    (u_2+\tau)^3}}\,du_1du_2.  
\ee
Section \ref{VWP2} will discuss that this
modular anomaly for $SU(3)$ VW theory may be traded for a holomorphic
anomaly, similarly to the discussion for gauge group $SU(2)$. The
detailed structure is such that the $SU(3)$ partition function is an
example of what is mathematically known as a mock modular form of 
depth two. Integrals of the form (\ref{itintegral}) (but involving
weight $\frac{3}{2}$ modular forms) have recently also appeared in
the study of vertex operator algebras and quantum modular forms
\cite{bringmann2017higher}. Similar integrals for
integer weight modular forms have been
studied in \cite{Manin:2005}, which have recently found applications
in the context of Feynman amplitudes \cite{Adams:2017ejb}.
 
The derivation of the new results for VW partition functions is based on the holomorphic partition functions, which
were derived in earlier work \cite{Manschot:2010nc,
  Manschot:2014cca}. Reference \cite{Manschot:2014cca} expressed the
partition function in terms of Appell functions of
signature $(2,2)$, which can in turn be related to indefinite theta
series whose associated lattice has signature $(2,2)$. To determine
the behavior of these functions under modular transformations, we
first trade the modular anomaly of the theta series for a holomorphic anomaly by 
adding specific subleading, non-holomorphic terms to the kernel of the
theta series. These terms were
determined in Reference \cite{Alexandrov:2016enp}, with the aid   
of twistorial techniques for D-brane instanton corrections in IIB
string theory \cite{Alexandrov:2012au, Alexandrov:2016tnf}. We will
relate these non-holomorphic terms to a period integral of a modular
form \cite{Alexandrov:2016enp, bringmann2017higher}. Modular
transformations of these integrals can be determined quite straightforwardly, from
which we in turn can deduce the modular properties of the holomorphic
partition function. 

The modular transformations of the $SU(3)$ partition function have a number of interesting consequences. First of all,
it confirms the proposed electric-magnetic duality of the VW theory. Given the
technical difficulties involved in the partition function such as
wall-crossing, and blow-up formula, this is quite
remarkable. We will demonstrate moreover in Section \ref{UNholan} that
for gauge group $U(3)$ the details work out neatly, such
that its ``holomorphic anomaly'' $D_NZ_N $ factors into the product $\sim Z_1
Z_2$. More precisely, for $N=1,2$ and 3, it is now verified
that the holomorphic anomaly $D_NZ_N$ satisfies:\footnote{We refer to
  Section \ref{UNholan} for the definition of $D_N$ and other details.}
\be
D_NZ_N=-\frac{3i}{16\sqrt{2}\pi y^{\frac{3}{2}}} \sum_{k=1}^{N-1} k(N-k)\,Z_k\, Z_{N-k},
\ee
with $y=\mathrm{Im}(\tau)$. This gives further
evidence for such a factorization to hold for generic
$N$ \cite{Minahan:1998vr, Alim:2010cf}, in which case the $Z_N$ would
involve $(N-1)$-dimensional iterated integrals of modular forms,
and consequently mock modular forms of depth $N-1$. These functions will equally
play a role for other four-manifolds with $b_2^+=1$ \cite{Alim:2010cf,
  Manschot:2011dj}.   

Mock modular forms of higher depth have appeared in a few other
instances, i.e. the open Gromov-Witten theory of elliptic orbifolds
\cite{bringmann2016higher} and quantum invariants of torus
knots \cite{zbMATH06587777}. It would be interesting to see whether
the modular anomaly also has a physical or mathematical interpretation in these
examples. Beyond these examples, higher depth mock modular forms may play a more general role in conformal field theory,
gauge theory and string theory. In fact, realizing $U(N)$ VW theory as
a bound state of $N$ M5-branes in M-theory does naturally
provide links with these subjects. The $N$ M5-branes wrapped on
$\IP^2$ lead to the VW theory we consider, but they can equally be
wrapped on other four-manifolds or on divisors in Calabi-Yau threefolds
\cite{Maldacena:1997de}. The corresponding partition functions
\cite{Gaiotto:2006wm, deBoer:2006vg, Denef:2007vg} can be quite
intricate \cite{Alexandrov:2016tnf, Manschot:2009ia, Manschot:2010xp} for non-Ricci
flat divisors. We hope that the results for $\IP^2$ will help to
understand the partition functions for these more complicated geometries. Finally, reducing the M5-brane degrees of freedom to
two-dimensions leads to a $(0,4)$ conformal field theory whose
partition function is expected to coincide with the partition function
of the M5-branes \cite{Maldacena:1997de}.  Partition functions of
M5-branes have been connected recently to the moonshine phenomenon \cite{Cheng:2017dlj}, which suggests that
one may also hope to find such a connection for the $SU(N)$ VW
partition functions with $N\geq 3$.

\subsection*{Structure of the paper}
The structure of the paper is as follows. Section \ref{mod123}
reviews the $SU(N)$ and $U(N)$ Vafa-Witten partition functions for
$N=1,2$, and discusses the new results for $N=3$. The following
sections derive the new results for $SU(3)$ VW theory. For a self-contained exposition, we
have included Section \ref{Mod_forms}, which is an introductory section on modular
forms and mock modular forms. Section
\ref{indeftheta} reviews Jacobi forms and aspects of
indefinite theta series for signatures $(n-1,1)$ and $(n-2,2)$, and
their completions following \cite{Zwegers-thesis,
  Alexandrov:2016tnf}. Section \ref{AppellLerch} applies this to
building blocks of the VW partition functions, so-called (generalized)
Appell-Lerch sums. We combine all the ingredients in Section
\ref{VWP2}, and determine the modularity of the partition
functions.

\section{Modularity of Vafa-Witten theory}
\label{mod123}
This section discusses the modularity of VW theory. We concentrate on
explaining the novel results and postpone the detailed derivation to later sections. Subsection
\ref{VWSU23} deals with the partition functions for gauge
group $U(N)$ with a fixed magnetic 't Hooft flux, such as the partition functions for $SU(2)$ and
$SU(3)$. Subsection \ref{UNholan} discusses the $U(N)$ partition functions
by including a sum over $U(1)$ fluxes. 

\subsection{Partition functions for gauge groups $SU(2)$ and $SU(3)$}
\label{VWSU23}
The VW twist of $\mathcal{N}=4$ supersymmetric
Yang-Mills theory with gauge group $SU(N)$ contains a commuting BRST-like operator,
$\mathcal{Q}$. For a suitable $W$, the topologically
twisted action $\mathcal{S}_{\rm twisted}$ of VW theory can be
expressed as a $\mathcal{Q}$-exact term $\left\{ \mathcal{Q},W
\right\}$, plus a term multiplying the complexified coupling constant
$\tau$ (\ref{tau}):
\be
\label{action}
\mathcal{S}_{\rm twisted}=\left\{ \mathcal{Q},W  \right\}-2\pi i \tau (n-\Delta),
\ee
where $n$ denotes the instanton number,
\be
n=\frac{1}{8\pi^2} \int_M \mathrm{Tr}\, F\wedge F,
\ee 
and $\Delta$ is a rational number arising due to $R^2$
couplings in the YM action on a curved four-manifold $M$
\cite{Harvey:1996ir, Bachas:1999um}. To be precise, $\Delta$ equals
$N\chi(M)/24$ with $\chi(M)$ the Euler characteristic of $M$.

The path integral of VW theory localizes on a set of
equations whose solutions are graded by the instanton number
$n$ \cite{Vafa:1994tf}. Since $\mathcal{Q}-$exact terms decouple from the partition
function (to first approximation), the path integral includes a holomorphic $q$-series,   
\be    
\label{hN}
h_{N,0}(\tau)=\sum_{n\geq 0} c_N(n)\, q^{n-\Delta},  
\ee
where $q=e^{2\pi i \tau}$. The coefficient $c_N(n)$ is a topological invariant of 
the space $\cM_{N,n}$ of solutions to the VW equations with instanton number
$n$. In fact, $c_N(n)$ is the Euler
characteristic of $\cM_{N,n}$, $\chi(\cM_{N,n})$, for relative prime
$(N,n)$. We will specialize the four-manifold $M$ in the following to the complex projective
plane $\IP^2$. A vanishing theorem holds for this four-manifold, 
which has the consequence that the VW equations reduce to the
self-duality equation, $F=-*F$, or more generally the
Hermitian-Yang-Mills equations \cite{Vafa:1994tf}. The $c_N(n)$ is therefore a (weighted) Euler
characteristic of the moduli space of instantons on $\IP^2$ with
instanton number $n$ \cite{Vafa:1994tf}. In mathematical terminology,
the $c_N(n)$ is a Donaldson-Thomas type invariant of the moduli space of
semi-stable coherent sheaves on $\IP^2$.    

Before considering the partition function of the $SU(N)$ theories, it
is useful to make a few comments on the more general class of theories with gauge
group $U(N)$ and fixed magnetic 't Hooft flux $\frac{i}{2\pi} \mathrm{Tr}\, F
= \mu \in H^2(\mathbb{P}^2,\IZ_N)\cong \IZ_N$ \cite{Vafa:1994tf,
  tHooft:1979rtg, Girardello:1995gf}. The instanton number $n$ for a
generic flux $\mu$ is not
an integer for $\IP^2$, but takes values in $\IZ+\frac{\mu}{2}$. $S$-duality maps the
partition function $h_{N,\mu}$ of this theory to linear combinations
of the $h_{N,\nu}$, $\nu=1,\dots, N$. Their {\it expected} modular transformations under the generators $S$ and $T$ of
$SL(2,\mathbb{Z})$ are \cite{Vafa:1994tf}:
\be
\label{htrafos}
\begin{split}
S:&\qquad h_{N,\mu}\!\left(-\frac{1}{\tau} \right)=\frac{1}{\sqrt{N}}
(-i\tau)^{-\frac{3}{2}} (-1)^{N-1}\sum_{\nu \mod N} e^{-2\pi i
  \frac{\mu\nu}{N}} h_{N,\nu}(\tau),\\
T:&\qquad h_{N,\mu}(\tau+1)=e^{2\pi i(-\frac{N}{4}+\frac{1}{2N}(\mu+N/2)^2) } h_{N,\mu}(\tau).
\end{split}
\ee
These transformations follow most easily from the self-duality of
$U(N)$, which will be discussed in some more detail in Subsection \ref{UNholan}.
 
The partition functions $h_{N,\mu}$ are the building blocks of a family of
$SU(N)$ theories, namely the set of theories with gauge groups
$(SU(N)/\IZ_k)_{n \mod k}$ where $k$ divides $N$. The subscript $n$ labels different choices of
allowed line operators in the theory \cite{Gaiotto:2010be, Aharony:2013hda}. The partition function $h_{N,0}$ equals the one for $SU(N)$, while
those for the other groups are linear combinations of the $h_{N,\mu}$,
$\mu=0,\dots,N-1$. We deduce from Equation (\ref{htrafos}) that the $SU(N)$
theory is mapped to itself by $T$ and $ST^{N}S$ for $N$
odd, and $T$ and $ST^{2N}S$ for $N$ even. These are respectively the generators of the congruence subgroups
$\Gamma_0(N)$  and $\Gamma_0(2N)$.\footnote{The difference
  between the   congruence subgroups for even and odd $N$ is a consequence
  of the fact that $\mathbb{P}^2$ is not a spin manifold. The duality
  group for $SU(N)$ is $\Gamma_0(N)$ on a spin manifold for all $N$.}

The $h_{N,\mu}$ can in principle be evaluated for arbitrary $(N,\mu)$ \cite{Manschot:2011dj, Manschot:2010nc, Manschot:2014cca, Yoshioka:1994, Yoshioka:1995}. They take the form
\be
\label{formhNmu}
h_{N,\mu}=\frac{f_{N,\mu}}{\eta^{3N}},
\ee
where $\eta$ is the Dedekind eta function defined in Equation
(\ref{eta}), which is a modular form of weight $\frac{1}{2}$. The
numerator, $f_{N,\mu}$, can be thought of as the contribution to the
partition function from solution spaces of smooth instantons, while
$\eta^{-3N}$ is due to boundary components where instantons become point-like. 

Before describing the novel function appearing for $SU(3)$, let us start by reviewing the functions for $U(1)$ and
$SU(2)$. They are given by Equation (\ref{formhNmu}) with $f_{1,0}=1$ for
$U(1)$ \cite{Vafa:1994tf,
  Gottsche:1990}, while for $SU(2)$, $f_{2,0}=3G_0$ \cite{Vafa:1994tf, Yoshioka:1995, Klyachko:1991}
where $G_0$ is the generating function of Hurwitz class numbers \cite{Zagier:1975},
\be
\label{G0}
\begin{split}
G_0(\tau)&=\sum_{n\geq 0} H(4n)\,q^{n}\\
&=-\frac{1}{12}+\frac{1}{2} q+q^2+\frac{4}{3}q^3+\frac{3}{2}q^4+O(q^5).
\end{split}
\ee
More details of this function are given in Section \ref{Gclass}.
Curiously, it does not transform as a modular form under
$\Gamma_0(4)$. While $G_0$ is invariant under $T$, it
transforms under the second generator $ST^4S$ of $\Gamma_0(4)$ as 
\be   
\label{G0trafo}
\begin{split} 
G_0\!\left(\frac{- \tau}{4\tau-1}\right)=i (4\tau-1)^{\frac{3}{2}}
\left( G_0(\tau)+\frac{i}{4\sqrt{2}\pi}\int_{-\frac{1}{4}}^{i\infty}
  \frac{\Theta_0(u)}{(-i(\tau+u))^{\frac{3}{2}}}du \right),
\end{split}
\ee
where $\Theta_\alpha$ is the theta series 
\be
\label{Theta_alpha} 
\Theta_\alpha=\sum_{k\in \IZ+\alpha} q^{k^2}.
\ee
We notice that the transformation (\ref{G0trafo}) differs from the
usual transformation of a modular form, by a shift of a period
integral over $\Theta_0$. Such integrals were introduced and studied by Eichler \cite{Eichler1957} and
Shimura \cite{goroshimura1959}.

The transformation (\ref{G0trafo}) is a non-trivial confirmation of
$S$-duality, since the transformation reproduces $G_0$. However the
anomalous shift by a period integral requires explanation. A proposed resolution is that the partition
function includes a non-holomorphic part besides the holomorphic
$q$-series \cite{Vafa:1994tf, Minahan:1998vr}. If we define
\be
\label{hatf0}
\widehat f_{2,0}(\tau,\bar \tau)=f_{2,0}(\tau)-\frac{3i}{4\sqrt{2}\pi} \int_{-\bar \tau}^{i\infty}\frac{\Theta_0(u)}{(-i(\tau+u))^{\frac{3}{2}}}du,
\ee
then it transforms as a modular form of weight $\frac{3}{2}$ under 
$\Gamma_0(4)$. One may question the legitimacy of the
addition of the period integral, since the form of the action
(\ref{action}) suggests that the partition function is holomorphic in
$\tau$. However in the related context of topological string theory,
$\mathcal{Q}-$exact terms may contribute to the partition function as
a consequence of the boundary of the moduli spaces. The partition
function could acquire in this way a non-holomorphic
dependence \cite{Bershadsky:1993cx,
  Bershadsky:1993ta}. It is conceivable that the non-holomorphic
contribution may be derived along these lines in VW theory as being due to reducible $SU(2)$ connections $A=\left( \begin{array}{cc} a & 0 
    \\ 0 & b \end{array}\right)$, where $a$ and $b$ are $U(1)$
connections with opposite 't Hooft fluxes \cite{Vafa:1994tf, Minahan:1998vr}. Indeed, the non-holomorphic term
of $\widehat f_{2,0}$ is multiplied in $\widehat h_{2,0}=\widehat f_{2,0}/\eta^6$ by
$\eta^{-6}=h_{1}^2$. When we discuss the holomorphic
anomaly for the $SU(3)$ partition function below, and for $U(2)$ and $U(3)$ in the next subsection, we
will find more evidence for this interpretation. 

While not in VW theory, the non-holomorphic contribution is understood
in a physical context, namely IIB string theory. The   
$\widehat h_{N,\mu}$ appear  in D3-brane instanton corrections to the
hypermultiplet geometry \cite{Alexandrov:2012au} in a suitably chosen
compactification of IIB string theory.  The non-holomorphic period
integral appears in the form of a twistor integral
\cite{Alexandrov:2012au}. The form (\ref{hatf0}) is characteristic for functions known as mock modular forms in
mathematics \cite{Zwegers-thesis, MR2605321}. The function $\Theta_0$ is
known as the ``shadow'' of $f_{2,0}$ in this context. Section
\ref{MMF} will discuss mock modular forms in more detail. 
  
The main result of the present paper is the derivation of the modular properties of the
$f_{3,\mu}$, $\mu=0,1$, which is again of the form given in Equation (\ref{formhNmu}).
The first terms of $f_{3,0}$ are \cite{Manschot:2014cca,
  Manschot:2011ym}
\be      
\begin{split} 
f_{3,0}(\tau)&=\frac{1}{9}-q+3\,q^2+17\,q^3+41\,q^4+78\,q^5+120\,q^6+O(q^7).
\end{split}
\ee
Appendix \ref{Coeff_f3} gives an explicit expression for its $q$-series
 and lists the first 30 coefficients, which suggest that their growth is polynomial.
The function is clearly invariant under the generator $T\in
SL(2,\IZ)$. We will show in Section \ref{VWP2} that $f_{3,0}$ transforms under the other generator of
$\Gamma_0(3)$, $ST^3S$, as: 
\be
\label{f30shift}
f_{3,0}\!\left(\frac{-\tau}{3\tau-1} \right)=-(3\tau-1)^3\left(
  f_{3,0}(\tau) -\frac{i}{\pi} \left(
  \frac{3}{2}\right)^{\frac{3}{2}} \int_{-\frac{1}{3}}^{i\infty}  \frac{\sum_{\mu={0,1}}\widehat f_{2,\mu}(\tau,-u)\,\Theta_{\frac{\mu}{2}}(3u)}{(-i(\tau+u))^{\frac{3}{2}}}du\right).
\ee
Note that the completions $\widehat f_{2,\mu}$ appear in the integrand
of the period integral. Thus even the holomorphic function $f_{3,0}$
``knows'' about the non-holomorphic completion $\widehat f_{2,\mu}$. It is quite remarkable that  VW theory and the mathematical theory of
invariants of moduli spaces leads to $q$-series with such rich
properties. Moreover, since $\widehat f_{2,0}$ (\ref{hatf0}) involves
an integral over a theta series, we realize that the rhs of Equation
(\ref{f30shift}) includes an interated integral over theta series. As
will be explained in more detail in Section \ref{MMF}, $f_{3,0}$ is an
example of a mock modular form of depth two, since its shadow contains a
mock modular form (of depth one).

Similarly to the case of $SU(2)$, we can trade the modular anomaly by a
holomorphic anomaly by adding specific non-holomorphic
terms to $f_{3,0}$. Section \ref{secU3} will demonstrate that the completion $\widehat f_{3,0}$ is given by:
\be
\label{hatf30}
\begin{split}
\widehat f_{3,0}(\tau,\bar \tau)& =f_{3,0}(\tau)-\frac{i}{\pi} \left(
  \frac{3}{2}\right)^{\frac{3}{2}} \int_{-\bar \tau}^{i\infty}
\frac{\sum_{\mu={0,1}} \widehat f_{2,\mu}(\tau,-u)\,\Theta_{\frac{\mu}{2}}(3u)}{(-i(\tau+u))^{\frac{3}{2}}}du.
\end{split} 
\ee
The non-holomorphic term has again the structure reminiscent
of reducible connections. Moreover, substituting $\widehat f_{2,0}$ in this equation shows that $\widehat
f_{3,0}$ contains a (non-holomorphic) iterated period
integral over theta series. 

Before moving on to gauge group $U(N)$, let us discuss an experimental
observation. Note that we are free to add a vector-valued modular form of
weight three to $f_{3,\mu}$ with the same multiplier system, without changing the
completion. There is in fact one such a modular form given in Equation (\ref{TE6})
\cite{Manschot:2008zb}. In this way, we could for example cancel the
constant term of $f_{3,0}$ by subtracting $\frac{1}{9}\Theta_{E_6}$. The resulting function is 
in fact a mock modular cusp form of depth two. Interestingly, the
coefficients of the resulting function appear to be divisible
by 9. More explicitly we have, 
\be
\begin{split}
F&=\frac{1}{81}\Theta_{E_6}-\frac{1}{9}f_{3,0}\\
&=q+3\,q^2+7\,q^3+7\,q^4+18\,q^5+14\,q^6+23\,q^7+30\,q^8+O(q^9).
\end{split}
\ee
It would be interesting to explore why the coefficients of $f_{3,0}$
satisfy this congruence, and whether these coefficients have an independent
arithmetic interpretation.

\subsection{Gauge group $U(N)$ and the holomorphic anomaly}
\label{UNholan}
We have seen in the previous section, that the VW partition functions
for $SU(2)$ and $SU(3)$ transform as a modular form, once non-holomorphic terms
are added to the holomorphic generating series. We recall in this
subsection that this holomorphic anomaly fits a quite elegant
holomorphic anomaly equation for the $U(2)$ partition function
\cite{Minahan:1998vr}, and derive the holomorphic anomaly equation for
$U(3)$.

The partition function for gauge group $U(N)$ includes a sum over the
't Hooft fluxes (or first Chern classes) $c_1=\frac{i}{2\pi }\mathrm{Tr}\, F\in
H^2(\IP^2,\IZ)$. Fortunately, a symmetry allows us to include this sum
relatively easily. Namely, addition to the field strength $F$ of 
$$-2\pi i\, \omega\, k\, {\bf 1}_N,$$ 
with $\omega$ the K\"ahler form, $k\in \IZ$ and ${\bf 1}_N$ the
$N$-dimensional identity matrix, induces an isomorphism between the instanton moduli spaces $\cM_\gamma$ and
$\cM_{\gamma'}$. The transformed vector $\gamma'=(N,c_1',n')$ is given by 
\be
\begin{split}
& c_1'= c_1 + k N\, \omega,\\
& n'= n +\frac{1}{2} k^2N+\frac{k\,i}{2\pi}\mathrm{Tr} F,
\end{split}
\ee   
and as a result $h_{N,\mu}=h_{N,\mu+Nk}$. Due to this symmetry, the $U(N)$ partition function allows a theta-function decomposition
\be
\label{ZN} 
Z_N=\alpha_N\,\sum_{\mu\in \IZ_N} \widehat h_{N,\mu}\,\overline{\vartheta}_{N,\mu},
\ee
where the $\widehat h_{N,\mu}$ are appropriate completions of the $h_{N,\mu} $ (\ref{hN}), and we include a
dimensionless constant $\alpha_N$, which will prove useful later in
this section.\footnote{Such constants are also familiar from
  Donaldson-Witten theory \cite{Moore:1997pc}.} Besides the equality $h_{N,\mu}=h_{N,\mu+Nk}$, the $h_{N,\mu}$ satisfy
the relation $h_{N,\mu}=h_{N,-\mu}$. There are thus only $\lfloor
\frac{N}{2}\rfloor +1$ independent $h_{N,\mu}$. As a result, the
apparently $N$-dimensional representation of $SL(2,\IZ)$ in Equation
(\ref{htrafos}), is in fact only  $\lfloor
\frac{N}{2}\rfloor +1$ dimensional.

 The theta series $\vartheta_{N,\mu}$ in Equation (\ref{ZN}) captures the sum over $U(1)$
 fluxes, and is given by
\be
\label{vartheta}
\vartheta_{N,\mu}(\tau,\rho)=\sum_{k\in \mu+\frac{N}{2} +N\IZ} (-1)^{k}
q^{\frac{k^2}{2N}}e^{2\pi i \rho k }. 
\ee 
Note we added a fugacity $\rho$ for the 't Hooft flux. We discuss
these theta series in a bit more detail in Section \ref{secJacobi}. A few comments
are in order. To properly include fermions in the theory, the flux $k$ in (\ref{vartheta}) is shifted
by $N/2$ for $\IP^2$ such that $F$ is a spin$_c$ connection
\cite{Freed:1999vc}. Explicitly, we have
\be
k=\frac{i}{2\pi}\int_{H}  \mathrm{Tr}\, F\in\frac{N}{2} \int_{H} w_2 +\IZ=\frac{N}{2}+\IZ,
\ee
with $w_2$ the second Stiefel-Whitney class of $\IP^2$ and $H$ its
hyperplane. The phase $(-1)^{k}$ is a
consequence of integrating out massive modes of the fermions \cite{Witten:1995gf}.
 
The partition function $Z_N$ is conjectured to transform as a modular
form of mixed weight $(-\frac{3}{2},\frac{1}{2})$ under
$SL(2,\IZ)$. One can either arrive at this result from the
topologically twisted theory \cite{Vafa:1994tf}, or by reducing the degrees of freedom
to two dimensions using M5-branes \cite{Minahan:1998vr, Gaiotto:2006wm, deBoer:2006vg, Denef:2007vg} and using the
modularity of two-dimensional conformal field theory. More precisely, the 
expected transformation properties for $\IP^2$ are
\be
\begin{split}
&Z_N\!\left(-\frac{1}{\tau},\frac{\rho}{\tau}\right)=i^{-N} \tau^{-\frac{3}{2}} {\bar
  \tau}^{\frac{1}{2}} e^{\pi i\frac{N\rho^2}{\tau}} Z_N(\tau,\rho),\\
&Z_N(\tau+1,\rho)=i^{-N} Z_N(\tau,\rho). 
\end{split}
\ee
Combining these transformations with the transformations of
$\vartheta_{N,\mu}$ (\ref{varthetaST}), implies the transformations of
the $h_{N,\mu}$ (\ref{htrafos}). 

Let us now turn to the holomorphic anomaly of the functions $\widehat
h_{N,\mu}$, which was introduced to mitigate the modular anomaly of
$h_{N,\mu}$. To demonstrate the effect of the holomorphic anomaly, we
act with the wave operator
\be
D_N=\partial_{\bar \tau}-\frac{i}{4\pi N} \partial_{\bar \rho}^2,
\ee
on $Z_N$. The operator $D_N$ annihilates the $\vartheta_{N,\mu}$, and
receives therefore only contributions from the non-holomorphic terms
added to $h_{N,\mu}$. Working out the details for $N\leq 3$, we find
an interesting structure. 

Clearly, $D_1Z_1$ vanishes. Continuing with
$D_2Z_2$, we find using Equations (\ref{hatf0}) and (\ref{f2complete})
\be  
D_2Z_2=-\frac{3i\alpha_2}{16\pi y^{\frac{3}{2}}} \frac{1}{\eta^6} \left(\overline{\Theta}_0\,\overline{\vartheta}_{2,0}+\overline{\Theta}_{\frac{1}{2}}\,\overline{\vartheta}_{2,1}\right).
\ee
To write this more elegantly, we use the identity:
\be
\Theta_0\,\vartheta_{2,0}+\Theta_{\frac{1}{2}}\,\vartheta_{2,1}=\vartheta_{1,0}^2,
\ee
to arrive at \cite{Minahan:1998vr, Alim:2010cf}
\be
\label{D2Z2}
D_2Z_2=-\frac{3i}{16\pi y^\frac{3}{2}} \frac{\alpha_2}{\alpha_1^2}\, Z_1^2.
\ee
As mentioned before, the factorization of the anomaly into $Z_1^2$ is suggestive of an
explanation in terms of reducible connections, where the $U(2)$
connection is rather a $U(1)\times U(1)$ connection.

Using the new results for $U(3)$ with fixed 't Hooft flux, we can also determine $D_3Z_3$. Using Equations (\ref{hatf30}) and
(\ref{hatf31}), this becomes
\be 
\begin{split} 
&D_3Z_3=-\frac{i\alpha_3}{\pi} \frac{3^\frac{3}{2}}{8\pi
  y^{\frac{3}{2}}} \frac{1}{\eta^9} \\
&\times \sum_{\mu =0,1} \widehat f_{2,\mu}
  \left[ \overline{\Theta}_\frac{\mu}{2}(3\tau)\,\overline{\vartheta}_{3,0}(\tau,\rho)+\overline{\Theta}_{\frac{2+3\mu}{6}}(3\tau)\,\left(\overline{\vartheta}_{3,1}(\tau,\rho)+\overline{\vartheta}_{3,-1}(\tau,\rho)\right)
\right].  \\
\end{split}
\ee
Similarly to $U(2)$, we use an identity for the products of theta series:
\be
\begin{split}
&\Theta_\frac{\mu}{2}(3\tau)\,\vartheta_{3,0}(\tau,\rho)+\Theta_{\frac{2+3\mu}{6}}(3\tau)\,\left(\vartheta_{3,1}(\tau,\rho)+\vartheta_{3,-1}(\tau,\rho)\right)\\
&=\vartheta_{1,0}(\tau,\rho)\,\vartheta_{2,\mu}(\tau,\rho).
\end{split}
\ee
After substitution of this expression, we can express the rhs in terms
of $Z_1$ and $Z_2$:
\be 
\label{D3Z3}
D_3Z_3=-\frac{i3\sqrt{3}}{8 \pi y^{\frac{3}{2}}} \frac{\alpha_3}{\alpha_1\alpha_2} Z_1 Z_2. 
\ee
Again we recognize the factorization of the holomorphic anomaly, which
is in this case suggestive of reducible connections $U(1)\times U(2) \subset U(3)$. 

Besides for the four-manifold $\mathbb{P}^2$, such a
structure was also found for the
rational elliptic surface, $\frac{1}{2}$K3 \cite[Equation
(3.18)]{Minahan:1998vr}, where it is related to the holomorphic
anomaly of topological strings by $T$-duality. The general proposed
form for the anomaly $D_NZ_N$ for four-manifolds
$M$ with $b_2^+=1$ is \cite{Minahan:1998vr, Alim:2010cf}
\be
\label{DNZN}
D_N Z_N=C_M(y)\, \sum_{k=1}^{N-1} k(N-k)\, Z_k  Z_{N-k}. 
\ee 
Note that the factor $k(N-k)$ equals the number of matrix
elements of a $U(N)$ connection, which vanish if the connections is
reducible to $U(k)\times U(N-k)\in U(N)$. Equation (\ref{DNZN}) is
indeed confirmed by Equations  (\ref{D2Z2}) and (\ref{D3Z3}), if we
take $\alpha_N \sim\frac{1}{\sqrt{N}}$ for $\IP^2$. It would be
interesting to derive this factor within VW theory. It is promising
that the structure (\ref{ZN}) with $\alpha_N\sim1/\sqrt{N}$ does
occur naturally in the hypermultiplet geometry of
IIB string theory. See Equation (4.14) in Reference
\cite{Alexandrov:2012au}.\footnote{Note after specializing the
  generic case discussed in \cite{Alexandrov:2012au} to $U(N)$ VW
  theory on $\IP^2$, the prefactor $1/\sqrt{\bfp\cdot \bft^2}$ in
  \cite[Equation (4.14)]{Alexandrov:2012au} becomes proportional to
  $1/\sqrt{N}$.} 

A natural question is the generalization of Equation (\ref{DNZN}) to
other four-manifolds with $b_2^+=1$. If in addition $b_2>1$, the situation is
more complicated due to the effect of wall-crossing. For Hirzebruch surfaces, which are examples of such
four-manifolds, it was found in Reference \cite{Manschot:2011dj} for gauge group
$U(2)$ that additional terms are
present on the rhs of Equation (\ref{DNZN}) for a generic choice of
metric, while the additional terms may vanish for special metrics. This
happens most notably if the period point of the metric is proportional
to the anti-canonical class $-K_M$.

\section{Modular forms and mock modular forms}
\label{Mod_forms}
We begin in this section the derivation of the results described
in Section \ref{mod123}. We recall a number of aspects of modular forms, which
will be useful in the following sections. The discussion is 
largely based on examples, which will return in the following
sections. For more details on modular forms, the reader may consult
one of the many textbooks on the subject, such as \cite{apostol, Diamond:2005}.

\subsection{Modular groups}
The modular group $SL(2,\IZ)$  is defined by: 
\be
SL(2,\IZ)=\left\{ \left. \left(\begin{array}{cc} a & b \\ c &
      d \end{array}\right) \right\vert a,b,c,d\in \IZ;\,\, ad-bc=1 \right\}. 
\ee
This group is generated by two elements
$S=\left(\begin{array}{cc} 0 & -1 \\ 1 & 0 \end{array} \right)$ and $T=\left(\begin{array}{cc} 1 & 1 \\ 0 & 1 \end{array} \right)$. 

Two congruence subgroups of $SL(2,\IZ)$ are relevant for us. The congruence subgroup $\Gamma_0(n)\in SL(2,\IZ)$ is defined as:
\be
\Gamma_0(n)=\left\{\left. \left(\begin{array}{cc} a & b \\ c & d \end{array}\right)\in
    SL(2,\IZ) \right\vert c=0\mod n \right\}. 
\ee
Its generators are $ST^nS$ and $T$.  Finally, the congruence subgroup $\Gamma(n)\in SL(2,\IZ)$ is defined as
\be
\Gamma(n)=\left\{\left. \left(\begin{array}{cc} a & b \\ c & d \end{array}\right)\in
    SL(2,\IZ) \right\vert \left(\begin{array}{cc} a & b \\ c &
      d \end{array}\right)=\left(\begin{array}{cc} 1 & 0 \\ 0 &
      1 \end{array}\right) \mod n \right\}. 
\ee

\subsection{Modular forms}
A modular form $f:\mathbb{H}\to \mathbb{C}$ of weight $k$ for $SL(2,\IZ)$ is a
function which satisfies 
\be
\label{moddef}
f\!\left( \frac{a\tau+b}{c\tau+d}\right)=(c\tau+d)^kf(\tau),\quad
\mathrm{for\,\, all}
\quad \left(\begin{array}{cc} a & b \\ c & d\end{array}
\right)\in SL(2,\mathbb{Z}).
\ee 
We denote the space of holomorphic modular forms of weight $k$ with
subexponential growth for $\tau\to i\infty$ by $M_k(SL(2,\IZ))$. One
of the powerful features of modular forms is that the space $M_k$ is
finite dimensional for fixed $k$. We can similarly define modular forms
for congruence subgroups such as $\Gamma_0(n)$ and $\Gamma(n)$
introduced above. We can further
generalize Equation (\ref{moddef}) to include modular forms with a
multiplier system. These are functions which transform as in Equation (\ref{moddef}),
except that the right-hand side is multiplied by a phase $\varepsilon(\gamma)$ for each $\gamma\in
SL(2,\IZ)$. 

If the weight $k$ is half-integral, such a
multiplier system is in fact required to have any non-trivial holomorphic
functions satisfying Equation (\ref{moddef}). For example, the Dedekind eta function $\eta$, defined by
\be
\label{eta}
\eta(\tau)=q^{\frac{1}{24}}\prod_{n=1}^\infty (1-q^n),
\ee
transforms under the generators $S$ and $T$, with the phases
$\varepsilon(S)=e^{-\frac{2\pi i}{8}}$ and $\varepsilon(T)=e^{\frac{2\pi i}{24}}$:
\be
\label{Dedekind}
\begin{split}
&S:\qquad \eta\!\left(-\frac{1}{\tau}\right)=\sqrt{-i\tau}\,\eta(\tau), \\
&T:\qquad \eta(\tau+1)=e^{\frac{2\pi i}{24}} \eta(\tau).
\end{split}
\ee

One of the key techniques to construct modular forms is as a theta
series, which are functions which involve a sum over a
lattice.\footnote{Another widely used technique is the sum over modular images,
  as for example in Eisenstein and Poincar\'e series.} Application of Poisson resummation allows to determine the
modular properties of such series relatively easily. The simplest example has associated lattice
$\IZ$, and is defined as $\Theta_0=\sum_{k\in \IZ}q^{k^2}$. This is a modular form
for the group $\Gamma_0(4)$ with a multiplier system. Its transformation for an arbitrary element of this group is: 
\be
\label{Theta0trafos}
\Theta_0\!\left(\frac{a\tau+b}{c\tau+d}\right)=(c\tau+d)^{\frac{1}{2}}
\left( \frac{c}{d}\right)\varepsilon_d\, \Theta_0(\tau),\quad
\mathrm{for\,\, all}
\quad\left(\begin{array}{cc} a & b \\ c
    & d \end{array}\right)\in \Gamma_0(4),
\ee
where $\left( \frac{c}{d}\right)\in \pm 1$ is the Jacobi symbol and 
\be
\varepsilon_d=\left\{ \begin{array}{ll} 1, & d=1\mod 4, \\ i, & d=3 \mod
  4. \end{array}\right.  
\ee

We end this section with a brief discussion on theta series for
higher-dimensional lattices. We consider first the series $b_{3,j}$,
whose corresponding lattice is the $\mathrm{A}_2$ root
lattice:\footnote{To notation $b_{3,j}$ is chosen to match with previous work.
See for example References \cite{Manschot:2014cca, BMR2015}.} 
\be
\label{ThetaA2}
b_{3,j}(\tau):=\sum_{k_1,k_2\in \mathbb{Z}+\frac{j}{3}} q^{k_1^2+k_2^2+k_1k_2},  \\
\ee
These functions transform under
$\Gamma(3)$ for $j=0,1$, and form in fact a vector-valued
representation of $SL(2,\IZ)$. The three functions
$b_{3,j}$ with $j\in\{-1,0,1\}$ form a three-dimensional
representation of $SL(2,\IZ)$. For the generators $S$ and $T$, they
transform as
\begin{eqnarray}
&&S:\qquad b_{3,j}\!\left(-\frac{1}{\tau}\right)=
-\frac{i \tau}{\sqrt{3}} \sum_{\ell \mod 3} e^{-2\pi i j\ell/3} \,b_{3,\ell}(\tau),\\
&&T:\qquad b_{3,j}(\tau+1)=e^{2\pi i j^2/3}\,b_{3,j}(\tau).
\end{eqnarray}
In particular, $b_{3,0}$ is a modular form of weight 1 for the congruence subgroup $\Gamma_0(3)$ with multiplier
$\left(\frac{d}{3}\right)$ \cite{Borwein1994}. 

As a final example, we introduce the theta series $\Theta_{E_6}$,
whose associated lattice is the six-dimensional root lattice
$\Lambda_{E_6}$ of $E_6$. It is explicitly given as
\be
\label{TE6}
\begin{split}
\Theta_{E_6}(\tau)&= \sum_{k\in \Lambda_{E_6}} q^{Q(k)/2} \\
& = 1+ 72 q + 270q^2+720 q^3+936 q^4+\dots, 
\end{split} 
\ee
where $Q(k)$ is here the quadratic form of the $E_6$ root lattice.
The function $\Theta_{E_6}$ transforms as a modular form of
$\Gamma_0(3)$ and its weight is 3. In terms of the $b_{3,j}$ introduced above, we have the identity
\be
\label{TE6}
\Theta_{E_6}=b_{3,0}^3+2b_{3,1}^3.
\ee
This function forms together with $3b_{3,0}b_{3,1}^2$ a
two-dimensional representation of $SL(2,\IZ)$. After dividing these
functions by $\eta^9$, they transform precisely as the $\widehat
h_{3,\mu}$ in Equation (\ref{htrafos}). It can be shown that $\Theta_{E_6}$ is the only 
holomorphic modular form with this property \cite{Manschot:2008zb}.   

\subsection{Mock modular forms}
\label{MMF}

The previous subsection provided examples of holomorphic modular forms. The spaces of such functions are finite
dimensional once the weight and multiplier system are fixed. However,
as discussed in the previous section, these spaces are not rich enough
to capture the VW partition functions for four-manifolds with $b_2^+=1$. We have seen for gauge group
$SU(2)$, that we should consider
functions, which can be expressed as a holomorphic $q$-series plus a non-holomorphic period
integral. Such functions are known as mock modular forms, which have
received much interest in recent years following
\cite{Zwegers-thesis, MR2605321}. We will discuss the main aspects of these functions in this subsection. See
Reference \cite{BFOR2018} for a recent comprehensive text book on the subject.
 
We restrict the discussion for simplicity to the full
modular group $SL(2,\IZ)$. The generalization to congruent subgroups,
such as $\Gamma_0(3)$ and $\Gamma_0(4)$ is straightforward. To introduce the notion of a mock modular form, let us first introduce
the so-called ``shadow map'' \cite{MR2605321}. The argument of this map is a function
$g:\mathbb{H}\times \mathbb{\overline H}\to \mathbb{C}$ with mixed
weight $(\ell,2-k+\ell)$, i.e. $g$ transforms under $SL(2,\IZ)$ as
\be
\label{gtrafo}
g\!\left( \frac{a\tau+b}{c\tau+d}, \frac{a \sigma+b}{c
    \sigma+d}\right)=(c\tau+d)^{\ell} (c\sigma+d)^{2-k+\ell} g(\tau,\sigma).
\ee
The shadow map sends $g$ to the non-holomorphic
period integral $g^*$, defined as
\be
\label{shadowmap}
g^*(\tau,\bar \tau)=-i(1/2)^{k-1} \int_{-\bar \tau}^{i\infty} \frac{
g(\tau,-v)}{(-i(v+\tau))^{k-\ell}}\, dv. 
\ee 
This function transforms under an element of $SL(2,\mathbb{Z})$ almost
as a modular form of weight $k$, but the transformation includes a
shift by a holomorphic period integral:
\be 
\label{g*trafo}
g^*\!\left( \frac{a\tau+b}{c\tau+d}, \frac{a \bar \tau+b}{c
    \bar \tau+d}\right)=(c\tau+d)^k \left( g^*(\tau,\bar \tau) +i
  (1/2)^{k-1} \int_{\frac{d}{c}}^{i\infty}\frac{g(\tau,-v)}{(-i(\tau+v))^{k-\ell}}dv \right).
\ee
 
Using the shadow map, we can introduce the notions of completion, shadow and
of mock modular form. Let $h$ be a holomorphic $q$-series. Then we
call the sum,
\be
\label{hcompr}
\widehat h=h+g^{*},
\ee
the completion of $h$, if $\widehat h$ transforms as a modular form
for some weight $k$. In this case, we call the function $g$ the ``shadow'' of
$h$. Note that the shadow of
$h$ can be obtained from $\widehat h$ by acting with $y^{k-\ell}\partial_{\bar
  \tau}$, where $y=\mathrm{Im}(\tau)$. Of course, most $q$-series do not have a completion and shadow,
since modular transformations of an arbitrary $q$-series do not
transform complementary to the transformation of $g^*$ in Equation
(\ref{g*trafo}). 

A mock modular form $h$ is a holomorphic $q$-series, whose shadow $g$ is required to factor as $g=f_1
\overline{f}_2$, where $f_1$ is a holomorphic modular form of weight
$\ell$ and $f_2$ is a holomorphic modular form
of weight $2-k+\ell$. In other words, $g$ is an element of the tensor
space $ M_\ell \otimes\overline{M}_{2-k+\ell}$. When $\ell=0$ and
$f_1$ a constant, we say that the mock modular form $h$ is ``pure''
 \cite{Dabholkar:2012nd}, while when $f_1$ is a non-constant modular
form, $h$ is called a ``mixed'' mock modular form. The class number
generating function $G_0$ (\ref{G0}) is an example of a pure mock modular 
form for the congruence subgroup $\Gamma_0(4)$. This can be seen by comparing Equations (\ref{G0trafo}) and (\ref{g*trafo}), using that $\Theta_0(-\bar
\tau)$ equals the complex conjugate $\overline{\Theta_0(\tau)}$. The
class of mock modular forms can be expanded further, by allowing $g$
to be a sum over products, $\sum_{j} f_{1,j} \overline{f}_{2,j}$, with
weights $\ell_j$ and $2-k+\ell_j$ \cite{Dabholkar:2012nd}.  
 
The modular transformations of the mock modular forms discussed above
only involve one-dimensional period integrals. To include also functions
involving higher dimensional iterated integrals in the theory of mock
modular forms, the notion of   
``depth'' of a mock modular form is introduced
\cite{ZagierZwegers, Zwegerstalk}. The depth is a positive number, which gives information on the shadow $g$ of $h$. Mock modular forms of depth zero coincide with the familiar holomorphic
modular forms, whose shadow vanishes, while the mock modular forms of
the previous paragraph are said be of depth one. The depth is defined iteratively for $r>1$. To 
this end, let us denote the space of mock modular forms of
depth $r$ by $\mathbb{M}_k^{r}$, and the space of their
completions by $\mathbb{\widehat  M}_k^{r}$. We say that a mock modular form
has depth $r$ if its associated shadow $g$ is an element of the tensor space
$\mathbb{\widehat M}_{\ell}^{r-1}\otimes \overline{M}_{2-k+\ell}$ for
some $\ell$.\footnote{For simplicity, we assume here that $g$ factors as a product
  $f_1\overline{f}_2$. The generalization to a shadow $g=\sum_{j} f_{1,j}
  \overline{f}_{2,j}$ as mentioned above is straightforward.} We note that a mock modular form of depth $r$, involves
$r$-dimensional iterated integrals. 

We can also extend iteratively the definition of ``pure'' to
depth $>1$. To this end, let $g=\widehat f_1 \overline{f}_2$, with $\widehat f_1\in \mathbb{\widehat M}_{\ell}^{r-1}$
and $f_2\in  M_{2-k+\ell}$. Then for $r>1$, we say that $h$ is pure, if
$\widehat f_1$ is pure. We deduce from Equation (\ref{hatf30}) that
the VW partition function $f_{3,0}$ is
a (sum of two) pure mock modular form(s) of depth two with weight $k=3$ and
$\ell=\frac{3}{2}$.

\subsection{Generating function of Hurwitz class numbers}
\label{Gclass}
As mentioned in the previous subsection, the class number generating
function is an example of a pure mock modular form of weight $3/2$. We
will discuss this function in some more detail in this section. 
As the name suggests, the generating function $G$ of Hurwitz class numbers 
$H(n)$ is defined as \cite{Zagier:1975, hirzebruch:1976}:
\be
\label{gen_class}
\begin{split}
G(\tau)&=\sum_{n\geq 0} H(n)\,q^n\\
&=-\frac{1}{12} + \frac{1}{3}q^3+\frac{1}{2}q^4+q^7+\dots,
\end{split} 
\ee
where the $H(n)$ are the Hurwitz class numbers.\footnote{The Hurwitz
  class number $H(-D)$ is defined for $D<0$ as the number of
binary integral quadratic forms, $Q(x,y)=Ax^2+Bxy+Cy^2$, $A,B,C\in
\IZ$, with negative discriminant $0>D=B^2-4AC$, modulo the
action of $SL(2,\IZ)$, and weighted by the inverse of the order of the automorphism
group in $PSL(2,\IZ)$. We set furthermore $H(0)=-\frac{1}{12}$.}
 Note that the $H(n)$ vanish for $d=1,2 \mod 4$. Using results going back to Kronecker \cite{Kronecker:1859}, the function $G$ can
be explicitly written as the following $q$-series:
\be
G(\tau)=-\frac{1}{2\Theta_0(\tau+\frac{1}{2})}\sum_{n\in \IZ}
\frac{n(-1)^nq^{n^2}}{1+q^{2n}}-\frac{1}{12} \Theta_0(\tau)^3.
\ee 

One
way to view this function is as an Eisenstein series for
$\Gamma_0(4)$ of weight 3/2, from which one may understand the modular
anomaly of $G$. Since the
weight is $\leq 2$, the classical definition of the Eisenstein series is
divergent, and the sum requires a regularization.
Due to the regularization, the holomorphic $q-$series does not transform as
modular form, but involves a shift by a
period integral. This can be mitigated by the addition of a
non-holomorphic period integral. The completion $\widehat G$, defined by \cite{Zagier:1975}
\be
\label{Gcomplete}
\widehat G(\tau,\bar \tau)=G(\tau)-\frac{i}{8\sqrt{2}\pi}\int_{-\bar
  \tau}^{i\infty} \frac{\Theta_0(u)}{(-i(\tau+u))^{\frac{3}{2}}} du,
\ee
transforms as a modular form of weight $\frac{3}{2}$ under
$\Gamma_0(4)$. 

We conclude this subsection by noting that a vector-valued non-holomorphic modular form $\widehat G_\mu$, $\mu=0,1$, for $SL(2,\IZ)$ can be obtained from
$\widehat G$. These functions appear in the $SU(2)$ and $SO(3)$ VW
partition functions, given in Equations (\ref{G0}) and
(\ref{f2j}). The holomorphic parts $G_\mu$ are given by
\be
G_\mu(\tau)=\sum_{n\geq 0} H(4n-\mu)\,q^{n-\frac{\mu}{4}},\qquad \mu=0,1,
\ee
while the modular completions read
\be
\label{hatGmu}
\begin{split}
\widehat G_\mu(\tau,\bar \tau)&=G_\mu(\tau)-\frac{i}{4\sqrt{2}\pi} 
\int_{-\bar \tau}^{i\infty}
\frac{\Theta_{\frac{j}{2}}(u)}{(-i(\tau+u))^{\frac{3}{2}}}\, du. 
\end{split}
\ee
The completed functions $\widehat G_\mu$ form a two-dimensional Weil
representation of $SL(2,\IZ)$:
\be
\begin{split}
&S:\qquad \widehat G_\mu\!\left( -\frac{1}{\tau},-\frac{1}{\bar \tau}\right)=-\frac{1}{\sqrt{2}}(-i\tau)^{\frac{3}{2}}
\sum_{\nu=0,1} (-1)^{\mu \nu}\, \widehat G_{\nu}(\tau,\bar \tau), \\
&T:\qquad \widehat G_{\mu}(\tau+1,\bar \tau+1)=(-i)^{\mu}\, \widehat
G_\mu(\tau,\bar \tau). 
\end{split}
\ee

The modular properties of the non-holomorphic period integral in
Equation (\ref{hatGmu}) are easily established. One finds for the
$\Gamma_0(4)$ generator $ST^4S$:
\be 
\label{period_trafo}
\int_{\frac{\bar \tau}{4\bar \tau-1}}^{i\infty}
\frac{\Theta_{0}(u)}{(-i(\tfrac{-\tau}{4\tau-1}+u))^{\frac{3}{2}}}\,
du=i(4\tau-1)^{\frac{1}{2}} \int_{-\bar \tau}^{-\frac{1}{4}}
\frac{\Theta_{0}(w)}{(-i(\tau+w))^{\frac{3}{2}}}\,
dw,
\ee
after the change of variables $u=\frac{w}{4w+1}$ and using Equation
(\ref{Theta0trafos}) for the transformation of $\Theta_0$. Then using
$\int_{-\bar \tau}^{-\frac{1}{4}}=\int_{-\bar
  \tau}^{i\infty}-\int_{-\frac{1}{4}}^{i\infty}$, we confirm the
transformation of the holomorphic part $G_0$ as in Equation (\ref{G0trafo}). Note that the shift is in particular holomorphic, even though
the lhs of Equation (\ref{period_trafo}) is non-holomorphic. This
transformation of the period integral then implies 
the transformation of the holomorphic $q$-series as in Equation (\ref{G0trafo}).

\section{Jacobi forms and indefinite theta series}
\label{indeftheta}
The previous section introduced modular forms and mock modular
forms. A well-known generalization of the single variable modular forms are Jacobi forms,
which include a second elliptic variable. The elliptic variable is
naturally included in theta series with a positive definite lattice, while the larger class of Jacobi
forms has diverse applications in mathematics and
physics. Similarly to the mock modular forms in the previous section,
a ``mock'' variation on theta series exists. These are theta series
whose associated lattice is indefinite. The techniques developed for
indefinite theta series with signature $(n-1,1)$ and $(n-2,2)$, are
the main tool to determine the properties of the $U(3)$ VW partition function.
In fact, the completion of the class number generating function may
also be derived via this route. 

The outline of this section is as follow. We start with a brief review
of Jacobi forms. Subsection discusses indefinite theta series for
signatures $(n-1,1)$ and $(n-2,2)$, and the generalized error
functions appearing in their completions. Subsequent 
subsections discuss decompositions of the generalized error functions,
and how these can be related to (iterated)
period integrals.

\subsection{Jacobi forms}
\label{secJacobi}
This subsection provides a brief introduction to Jacobi forms, and provides a
few useful examples. See for more properties of these functions the textbooks
\cite{BFOR2018, MR781735}.
Jacobi forms are functions of two variables, the modular variable $\tau\in \IH$ and the elliptic
variable $z\in \IC$. A Jacobi form $\phi: \mathbb{H}\times
\mathbb{C}\to \mathbb{C}$ is characterized by its weight $k$ and index
$m$. A Jacobi form of $SL(2,\IZ)$ satisfies for modular transformations:
\be
\label{Jacobitrafo}
\phi\!\left(\frac{a\tau+b}{c\tau+d},\frac{z}{c\tau+d}
\right)=(c\tau+d)^k e^{2\pi i \frac{m c z^2}{c\tau+d}}\phi(\tau,z),
\ee
and is quasi-periodic under shifts of $z$:
\be
\phi(\tau,z+k\tau+\ell)=q^{-m k^2} w^{-2m k}\phi(\tau,z),
\ee
where $w=e^{2\pi iz}$. Similarly to modular forms, these definitions are
easily modified to include congruence subgroups. One may also consider
multiple elliptic variables, in which case the index $m$ becomes a matrix.

Jacobi forms with half-integer weight and
index exist, if we include additional phases on the
right-hand-side of Equation (\ref{Jacobitrafo}). Famous examples with weight
$\tfrac{1}{2}$ and index $\tfrac{1}{2}$ are the Jacobi theta
functions: 
\be
\begin{split}
\label{Jacobitheta}
\theta_1(\tau,z)&=i \sum_{r\in
  \mathbb{Z}+\frac12}(-1)^{r-\frac12}q^{r^2/2}e^{2\pi i
  rz}, \\ 
\theta_2(\tau,z)&= \sum_{r\in
  \mathbb{Z}+\frac12}q^{r^2/2}e^{2\pi i
  rz},\\
\theta_3(\tau,z)&= \sum_{n\in
  \mathbb{Z}}q^{n^2/2}e^{2\pi i
  n z}, \\
\theta_4(\tau,z)&= \sum_{n\in 
  \mathbb{Z}} (-1)^nq^{n^2/2}e^{2\pi i
  n z}. 
\end{split}
\ee
The function $\theta_1$ is a Jacobi form for $SL(2,\IZ)$, whereas
the other three transform under $\Gamma_0(4)$. We will sometimes suppress
the $\tau$-dependence of the Jacobi theta functions and other Jacobi
forms, thus $\theta_j(\tau,z)=\theta_j(z)$. 

We define the more general family of binary theta series
$\vartheta_{N,\mu}(\tau,z)$, which we encountered earlier in Equation
(\ref{vartheta}) in the $U(N)$ VW partition function,
\be
\vartheta_{N,\mu}(\tau,z)=\sum_{k\in \mu+\frac{N}{2} +N\IZ} (-1)^{k}
q^{\frac{k^2}{2N}}e^{2\pi i z k }.
\ee
These functions form an $N$-dimensional Weil representation of
$SL(2,\IZ)$, with weight $\frac{1}{2}$ and index $\frac{N}{2}$. The
transformations under the generators $S$ and $T$ are
\be
\label{varthetaST}
\begin{split}
&S:\qquad \vartheta_{N,\mu}\!\left(-\frac{1}{\tau},\frac{z}{\tau}\right)=\frac{1}{\sqrt{N}}(-i\tau)^\frac{1}{2}
e^{-\pi i \frac{N^2}{2}+\pi i \frac{N v^2}{\tau}} \sum_{\nu \mod N} e^{-2\pi i \frac{\mu \nu}{N}}\,\vartheta_{N,\nu}(\tau,z),\\
&T:\qquad
\vartheta_{N,\mu}(\tau+1,z)=e\!\left(\textstyle{\frac{1}{2N}(\mu+\frac{N}{2})^2
  }\right) \vartheta_{N,\mu}(\tau,z).
\end{split} 
\ee

Theta series associated to higher dimensional, positive definite lattices provide similarly
examples of Jacobi forms. For example, we can easily include
an elliptic variable $z$ in $b_{3,j}$ (\ref{ThetaA2}) encountered in the last
subsection, by defining:
\be
\label{b3j}
b_{3,j}(\tau,z):=\sum_{k_1,k_2\in \mathbb{Z}+\frac{j}{3}} e^{2\pi i z(k_1+2k_2)}q^{k_1^2+k_2^2+k_1k_2}. \non \\
\ee
Under the generators $S$ and $T$, these $b_{3,j}$ transform as
\be
\begin{split}
&S:\qquad b_{3,j}\!\left(-\frac{1}{\tau}, \frac{z}{\tau}\right)=
-\frac{i \tau}{\sqrt{3}} e^{2\pi i z^2/\tau}\sum_{\ell \mod 3} e^{-2\pi i j\ell/3} \,b_{3,\ell}(\tau,z),\\
&T:\qquad b_{3,j}(\tau+1,z)=e^{2\pi i j^2/3}\,b_{3,j}(\tau,z),
\end{split}
\ee
and they satisfy the quasi-periodicity relation:
\be
\label{b3period}
b_{3,j}(\tau,z+\lambda \tau+\mu)=q^{-\lambda^2}w^{-2\lambda} b_{3,j}(\tau,z).
\ee 
We deduce from these relations that $b_{3,0}$ is a Jacobi form of weight one and index one for the congruence subgroup $\Gamma_0(3)$ with multiplier
$\left(\frac{d}{3}\right)$ \cite{Borwein1994}. For later reference, we express $b_{3,0}$ in terms of the Jacobi theta series,
\begin{eqnarray}
\label{cubictheta}
b_{3,0}(\tau,z)&=&\theta_2(6\tau,3z)\,\theta_2(2\tau,z)+\theta_3(6\tau,3z)\,\theta_3(2\tau,z).
\end{eqnarray}

\subsection{Indefinite theta series}
In the previous section, we met a few examples of theta series, whose 
associated lattice is definite, and which are examples of Jacobi
forms. This section will consider theta series, whose associated
lattice is indefinite. The relation between such theta series and the
theta series for a definite lattice is similar to the relation between
mock modular forms and classical modular forms, which we discussed in
Section \ref{MMF}. In fact, specialization of elliptic variables of an
indefinite theta series leads to mock modular forms.

As mentioned above, indefinite theta series involve a sum over an
indefinite lattice $\Lambda$. Summing over all lattice points as for
example in Equation (\ref{TE6}) gives a divergent series due to the
indefiniteness of the lattice. To ensure convergence, we introduce a
non-trivial kernel multiplying
$q^{Q(k)/2}$. Convergence can easily be ensured in this way, for
example by summing only over positive definite lattice
vectors. However, such a regularization does in general not lead to modular
transformations. As we will discuss in more detail, modular
transformation properties can be obtained for specific non-holomorphic
kernels \cite{Zwegers-thesis, Alexandrov:2016enp, Vigneras:1977}.  We will restrict in this section to signatures $(n-1,1)$ and
$(n-2,2)$, which are sufficient to determine the modular properties of
the $U(3)$ VW theory. More general signatures are relevant for the
analysis of higher rank gauge groups. References \cite{Nazaroglu:2016lmr,
  westerholt2016indefinite, funke2017theta} may be useful for such an analysis. 

We assume in the following that the lattice $\Lambda$ is non-degenerate, with
signature $(r,s)$ and dimension $n=r+s$ and $s=1$ or 2. The bilinear form
corresponding to $\Lambda$ is denoted by $B$, and the quadratic form by
$Q$. Let furthermore $p$ be a characteristic vector of $\Lambda$. We then define the theta series $\Theta:\IH\times \IC^n\to \IC$
associated to $\Lambda$ by
\be
\label{Theta}
\Theta[\Phi]_\mu(\tau,z)=\sum_{k\in
  \Lambda+\mu+\frac{1}{2}p} (-1)^{B(k,p)}\, \Phi(k+b)\,
q^{Q(k)/2}e^{2\pi i B(z,k)},
\ee
where $b$ is $\Im(z)/y$. If the kernel is trivial, $\Phi=1$, the series is clearly divergent
for $s>0$. Let us now specialize $\Lambda$ to a Lorentzian
lattice, or equivalently $s=1$. See for a more comprehensive
treatment \cite{Zwegers-thesis}. The holomorphic kernel $\Phi_1$,
which is most relevant for us has its support on a  positive definite
subset of $\Lambda$. To define $\Phi_1$, let $\sgn(x)$ be defined by 
\be
\sgn(x)=\left\{ \begin{array}{cc}  -1, & x<0,\\ 0, & x=0,\\ -1, & x>0.  \end{array}\right.
\ee
and let $C$ and $C'\in \Lambda \otimes \mathbb{R}$ be two negative definite
vectors, $Q(C), Q(C')<0$, and which satisfy in addition $B(C,C')<0$. The kernel
$\Phi_1:\Lambda\otimes \IR\to \{0,\pm \frac{1}{2},\pm 1\}$ is then
defined as
\be
\label{Phi1}
\Phi_1(x)=\frac{1}{2}\left(\sgn(B(C,x))-\sgn(B(C',x))\right).
\ee
The conditions for $C$ and $C'$ ensure that the series is
convergent. However, $\Theta[\Phi_1]_\mu$ does not transform as
a modular or Jacobi form, which can be understood  from the fact that
the support of $\Phi_1$ is not a lattice. Alternatively, one can check
that $\Phi_1$ does not satisfy the conditions derived by Vign\'eras
\cite{Vigneras:1977} for $\Theta[\Phi_1]_\mu$ to transform as a modular form. 

However, Zwegers \cite{Zwegers-thesis} has demonstrated that a non-holomorphic modular form
$\widehat \Theta_\mu=\Theta _\mu[\widehat \Phi_1]$ can be obtained from
$\Theta_\mu[\Phi_1]$.\footnote{Alternatively, one can use the results of
  Vign\'eras \cite{Vigneras:1977} to prove that the kernel $\widehat
  \Phi_1$ does lead to a function transforming as a modular form.}
 To this end, one replaces $\sgn(x)$ in $\Phi_1$ by the non-holomorphic smooth function
$E_1(\sqrt{2y} x)$ with $y=\mathrm{Im}(\tau)$, where $E_1\in C^{\infty}$ equals the Gaussian
integral 
\be
\label{error1}
\begin{split}
E_1(u)&=2\int_0^u e^{-\pi t^2} dt\\
&=\int_\IR e^{-\pi (t-u)^2} \sgn(t)\,dt.
\end{split}
\ee
One may express $E_1$ as a reparametrization of the error function, $E_1(u)=\operatorname{Erf}(\sqrt{\pi} u)$.
The completed indefinite theta series $\widehat
\Theta_\mu$ then reads:
\be
\widehat \Theta_\mu(\tau,z)=\sum_{k\in
  \Lambda+\mu+\frac{1}{2}p} (-1)^{B(k,p)}\, \widehat \Phi_1(k+b)\,
q^{Q(k)/2}e^{2\pi i B(z,k)},
\ee
where the kernel $\widehat \Phi_1:\Lambda\otimes \IR\to \IR$ is given by
\be
\widehat \Phi_1(\sqrt{2y} x)=\tfrac{1}{2}E_1(\sqrt{2y}\, B(\underline
C,x))-\tfrac{1}{2}E_1(\sqrt{2y}\, B(\underline C',x)).
\ee
Here $\underline C$ denotes the normalization of $C$,
$C/\sqrt{-Q(C)}$. Note that in the limit, $y\to \infty$, $\widehat \Phi_1
\to \Phi_1$. Similarly, if we let $C$ approach a null-vector
$D$ ($Q(D)=0$), then $E_1(\sqrt{2y} B(C,x))\to \sgn(B(D,x))$. The
function $\widehat \Theta_\mu$ transforms with weight $n/2$, and
multiplier system specified by $\Lambda$. The explicit transformations are:
\be 
\label{Thetatrafos}
\begin{split}
&S:\qquad  \widehat
\Theta_\mu\!\left(-\frac{1}{\tau},\frac{z}{\tau}\right)=\frac{i^{\frac{s-r}{2}}}{\sqrt{|\Lambda^*/\Lambda|}}\,\tau^{\frac{n}{2}}
e^{\pi i\frac{Q(z)}{\tau}-\pi i \frac{Q(p)}{2}}\sum_{\nu\in
\Lambda^*/\Lambda} e^{-2\pi i B(\mu,\nu)}\,\widehat \Theta_{\nu}(\tau,z),
\\
&T: \qquad \widehat \Theta_\mu(\tau+1,z)=e^{\pi i Q(\mu+\frac{1}{2}p)}\,\widehat \Theta_\mu(\tau,z),
\end{split}
\ee
with $r-s=n-2$ in this case.

Next we consider a lattice $\Lambda$ with signature $(n-2,2)$. A
convergent theta series can also be constructed in this case,  by
restricting the lattice sum to positive definite vectors. To this end, we
introduce the kernel $\Phi_2:\Lambda\otimes \IR\to \{0,\pm \frac{1}{4},\pm
\frac{1}{2},\pm 1\}$,  which involves two pairs of 
negative definite vectors $(C_i, C_i')$ for $i=1,2$. See for
additional sufficient requirements on the $C_i$ and $C_i'$ \cite[Theorem 4.2]{Alexandrov:2016enp}\cite{westerholt2016indefinite}\cite{kudla2016theta}. Then we set
\be
\Phi_2(x)=\frac{1}{4}\prod_{i=1}^2 \left( \sgn(B(C_i,x))-\sgn(B(C_i',x))\right).
\ee
With the kernel $\Phi_2$, we can again construct a convergent theta
series $\Theta_\mu$ as in Equation (\ref{Theta}) for Lorentzian
signature. For the same reason as for signature $(n-1,1)$, $\Theta_\mu$
does not transform as a modular form in general, but we can derive a
non-holomorphic completion of $\Theta_\mu$ similarly to the discussion
for $s=1$. To this end,
each product of $\sgn$'s in $\Phi_2$ is replaced by a generalized error
function $E_2$ \cite{Alexandrov:2016enp}
\be 
\label{E2M}
E_2(\alpha;u_1,u_2)=\int_{\IR^2} du_1'du_2'\,e^{-\pi (u_1-u_1')^2-\pi
  (u_2-u_2')^2} \sgn(u_2')\,\sgn(u_1'+\alpha u_2').
\ee
For the product $\sgn(B(C_1,x))\,\sgn(B(C_2,x))$, the arguments
$\alpha$, $u_1$ and $u_2$ of $E_2$ are given by
\be
\begin{split}
&\alpha=-\frac{B(C_1,C_2)}{\sqrt{\Delta(C_1,C_2)}},\,\qquad
\Delta(C_1,C_2)=Q(C_1)\,Q(C_2)-B(C_1,C_2)^2,\\
&u_1=-\frac{B(C_{1\perp 2},x)}{\sqrt{-Q(C_{1\perp 2})}},\qquad
u_2=-\frac{B(C_2,x)}{\sqrt{-Q(C_2)}}, 
\end{split}
\ee
where $C_{1\perp 2}$ is the component of $C_1$ orthogonal to $C_2$. The modular
properties of $\widehat \Theta_\mu$ are now as in Equation (\ref{Thetatrafos}), with $r-s=n-4$.

\subsection{Generalized error functions and period integrals}
To analyze the difference between the two kernels $\Phi_\ell$
and $\widehat \Phi_\ell$ (for $\ell=1,2$), we discuss in some more
detail the error
function $E_1:\mathbb{R}\to [-1,1]$, and the generalized error function
$E_2:\mathbb{R}^2\to [-1,1]$. We will in particular relate these to iterated
period integrals.   

We start with $E_1(u)\in C^\infty$, which is anti-symmetric and interpolates monotonically from $-1$ for $u\to -\infty$
to $+1$ for $u\to +\infty$. Let us
express $E_1(u)$ as $\sgn(u)$ plus a remainder term $M_1(u)$:
\be
\label{E1M}
E_1(u)=\sgn(u)+M_1(u).
\ee
Clearly, the function $M_1$ is discontinuous, since $\sgn$ is
discontinuous and $E_1$ is smooth. It has various integral
representations. Using the first expression in Equation (\ref{error1}) for $E_1$, one may write $M_1$ as 
\be
\label{M1u}
M_1(u)=\left\{ \begin{array}{cc} -\sgn(u)\,\beta_{\frac{1}{2}}(u^2), & u\neq 0,\\ 0, & u=0.   \end{array} \right. 
\ee
with
\be
\label{betanu}
\beta_\nu(x)=\int_x^\infty u^{-\nu} e^{-\pi u} du.
\ee

Using this expression, we may easily write $M_1(u)$ as a period integral: 
\be
\label{M1period}
M_1(u)=\frac{iu}{\sqrt{2y}}\, q^{\frac{u^2}{4y}} \int_{-\bar
  \tau}^{i\infty} \frac{e^{\frac{\pi i u^2
      w}{2y}}}{\sqrt{-i(w+\tau)}}dw,\qquad u\neq 0.
\ee
For later reference, we partially integrate the integral, such that
$uM_1$ can be expressed as
the form
\be
\label{uM1}
uM_1(u)=-\frac{1}{\pi} e^{-\pi
  u^2}-\frac{i \sqrt{2y}}{2\pi}\,q^{\frac{u^2}{4y}} \int_{-\bar
\tau}^{i\infty}\frac{e^{\frac{\pi i
    u^2w}{2y}}}{(-i(w+\tau))^{\frac{3}{2}}}dw.
\ee
This expression holds for all $u\in \IR$. 

We continue along the same lines with the generalized error function
$E_2$. We can express this function as a product of $\sgn$'s plus a remainder term \cite{Alexandrov:2016enp}:
\be 
\begin{split}
E_2(\alpha;u_1,u_2)&=\sign(u_2)\,\sign(u_1+\alpha u_2) +\sign(u_1)\,M_1(u_2)\\
& +\sign(u_2-\alpha
u_1)\,M_1\!\left(\frac{u_1+\alpha u_2}{\sqrt{1+\alpha^2}}\right)+ M_2(\alpha;u_1,u_2),  \\
\end{split}
\ee
where $M_1(u)$ is given by Equation (\ref{M1u}) as before. The function $M_2$ is
discontinuous across the loci where $u_1=0$ and $u_2-\alpha u_1=0$. To give the definition of $M_2$, we first
introduce the iterated integral $m_2$, 
\be
m_2(u_1,u_2)=2u_2\int_1^\infty dt \,e^{-\pi t^2 u_2^2} M_1(tu_1).
\ee
In terms of this function, $M_2$ is defined by
\be
M_2(\alpha;u_1,u_2)=\left\{ \begin{array}{ll}
    -m_2(u_1,u_2)-m_2\!\left( \frac{u_2-\alpha u_1}{\sqrt{1+\alpha^2}}, \frac{u_1+\alpha
    u_2}{\sqrt{1+\alpha^2}}\right), & u_1\neq 0,
u_2-\alpha u_1\neq 0, \\ [2ex]
-m_2\!\left( \frac{u_2}{\sqrt{1+\alpha^2}}, \frac{\alpha
    u_2}{\sqrt{1+\alpha^2}}\right), & u_1=0, u_2\neq 0,\\ [2ex]
-m_2(u_1,u_2), & u_1\neq 0, u_2-\alpha u_1=0, \\ [2ex]
\frac{2}{\pi} \arctan(\alpha),  & u_1=u_2=0.\end{array}\right. 
\ee  
Using the iterated integral $m_2$, we can in turn write $M_2$ as an iterated period integral
\cite{Alexandrov:2016enp, bringmann2017higher}. One finds for the various
domains of $u_1$ and $u_2$:
\begin{itemize}
\item[-] for $u_1\neq 0$ and $u_2-\alpha u_1\neq 0$:
\be
\label{E2period}
\begin{split}
&-\frac{u_1u_2}{2y}
q^{\frac{u_1^2}{4y}+\frac{u_2^2}{4y}}\int_{-\bar
  \tau}^{i\infty} dw_2 \int_{w_2}^{i\infty} dw_1 \frac{e^{\frac{\pi i
      u_1^2w_1}{2y}+\frac{\pi i
      u_2^2w_2}{2y}}}{\sqrt{-(w_1+\tau)(w_2+\tau)}}\\
&-\frac{(u_1+\alpha u_2)(u_2-\alpha u_1)}{2y (1+\alpha^2)}q^{\frac{u_1^2}{4y}+\frac{u_2^2}{4y}}\int_{-\bar
  \tau}^{i\infty} dw_2 \int_{w_2}^{i\infty} dw_1 \frac{e^{\frac{\pi i
      (u_2-\alpha u_1)^2w_1}{2(1+\alpha^2)y}+\frac{\pi i
      (u_1+\alpha
      u_2)^2w_2}{2(1+\alpha^2)y}}}{\sqrt{-(w_1+\tau)(w_2+\tau)}},
\end{split}
\ee
\item[-] for $u_1=0$, $u_2\neq 0$:
\be
\label{E2period2}
\begin{split}
&-\frac{\alpha u_2^2}{2y (1+\alpha^2)}q^{\frac{u_2^2}{4y}}\int_{-\bar
  \tau}^{i\infty} dw_2 \int_{w_2}^{i\infty} dw_1 \frac{e^{\frac{\pi i
      u_2^2w_1}{2(1+\alpha^2)y}+\frac{\pi i
      \alpha^2 u_2^2w_2}{2(1+\alpha^2)y}}}{\sqrt{-(w_1+\tau)(w_2+\tau)}},
\end{split}
\ee
\item[-] for $u_1\neq 0$, $u_1-\alpha u_2=0$:
\be 
\label{E2period3}
\begin{split}
&-\frac{u_1u_2}{2y}
q^{\frac{u_1^2}{4y}+\frac{u_2^2}{4y}}\int_{-\bar
  \tau}^{i\infty} dw_2 \int_{w_2}^{i\infty} dw_1 \frac{e^{\frac{\pi i
      u_1^2w_1}{2y}+\frac{\pi i
      u_2^2w_2}{2y}}}{\sqrt{-(w_1+\tau)(w_2+\tau)}},
\end{split}
\ee
\item[-] for $u_1=u_2=0$:
\be 
\label{u1u20}
\frac{2}{\pi} \arctan{\alpha}. 
\ee  
\end{itemize}

\section{Appell-Lerch sums and their completions}
\label{AppellLerch}
In this section, we will discuss a class of functions which are
closely related to the indefinite theta series of the previous
sections. The classical Appell-Lerch sums were introduced by Appell in 1886
\cite{Appell:1886} and also studied independently by Lerch
\cite{Lerch:1892}, in their study of doubly periodic  
functions. Over the years, various generalizations have appeared in
the mathematical \cite{Zwegers:2010, Polishchuk2001}  and physical
literature. For the latter, in
particular as partition functions in conformal field theories \cite{Semikhatov:2003uc} and topological
Yang-Mills theory \cite{Manschot:2014cca}. Similarly to the mock
modular forms and indefinite
theta series, these functions do not
transform as a modular form, but this can be mitigated by the addition
of a non-holomorphic, subleading term \cite{Zwegers-thesis, Alexandrov:2016enp}. 

Before discussing various examples, let us briefly present the general
form of the functions. As in Reference \cite{Manschot:2014cca}, we introduce a general Appell function in terms of an $m$-dimensional
positive definite lattice $\Lambda$, with associated bilinear form $B$ and
quadratic form $Q$. Let furthermore $\{ \bfm_j \}_{j=1,\dots,n}$ be
a set of $n$ vectors in the dual lattice $\Lambda^*$. The general
Appell function $A_{Q,{\bfm_i}}$ is then defined as:
\be
\label{AQ} 
A_{Q,{\bfm_i}}(\tau,\bfu,\bfv)=e^{2\pi i \ell(\bfu)}\sum_{k\in \Lambda}
\frac{q^{\frac{1}{2}Q(\bfk)+R}e^{2\pi i
    B(\bfv,\bfk)}}{\prod_{j=1}^{n}\left(1-q^{B(\bfm_j,\bfk)}e^{2\pi i
        u_j} \right)},
\ee
where $\bfu\in \IC^{n}$ and $\bfv\in \IC \times \Lambda$, and $\ell$ is a
linear function in $\bfu$. We will say that $A_{Q,{\bfm_i}}$ has
signature $(m,n)$. When we discuss the relation of these functions with
indefinite theta series later in this section, the signature of the
Appell function will coincide with the signature of the associated indefinite theta series.

We furthermore introduce the general Appell-Lerch sum $\mu_{Q,{\bfm_i}}$, which is almost identical to $A_{Q,{\bfm_i}}$,
except that it is divided by a theta series $\Theta_Q$ with
associated quadratic form $Q$,
\be
\mu_{Q,{\bfm_i}}(\tau,\bfu,\bfv)=\frac{A_{Q,{\bfm_i}}(\tau,\bfu,\bfv)}{\Theta_{Q}(\tau,\bfv)}.
\ee
We will focus in the following on examples, rather then discussing the
general functions introduced above. We will first recall the
classical Appell-Lerch sum and its completion, followed by a discussion
on the functions which appear in the partition
functions of $U(3)$ Yang-Mills theory.

\subsection{The classical Appell-Lerch sum}
\label{muA}
The classical Appell-Lerch sum $\mu(\tau,u,v):=\mu(u,v)$ is of signature $(1,1)$. Following Zwegers \cite{Zwegers-thesis}, we define
it as the ratio
\be
\label{muuv}
\mu(u,v)=\frac{A(u,v)}{\theta_1(v)},
\ee
with $\theta_1$ as in Equation (\ref{Jacobitheta}). The Appell function $A(\tau,u,v):=A(u,v)$  is defined as
\be
\label{Appell}
A(u,v)=e^{\pi i u}\sum_{n\in \mathbb{Z}}
\frac{(-1)^nq^{n(n+1)/2}e^{2\pi i n v}}{1-e^{2\pi i u} q^n}.
\ee
\\
\noi {\bf Completion of $\mu$}\\
As mentioned above, $\mu$ does not
transform as a (multi-variable) Jacobi form. However, a completion
$\widehat \mu(\tau,u,v)=\widehat \mu(u,v)$
can be defined, which does transform as Jacobi form. This function is
defined as \cite{Zwegers-thesis}:
\be
\label{Acomp}
\widehat \mu(u,v):=\mu(u,v)+\frac{i}{2}\,R(u-v),
\ee 
with  
\be
\label{Rutau}
R(\tau,u):=R(u)=\sum_{n\in \mathbb{Z}+\frac{1}{2}} \left(\,\sgn(n) - E_1\!\left((n+a)\sqrt{2y}\right)
  \right) (-1)^{n-\frac{1}{2}}e^{-2\pi i u n}q^{-n^2/2},
\ee
where $y=\Im(\tau)$, $a= \Im(u)/y$ and $E_1$ is the function as defined before
in Equation (\ref{error1}). Then $\widehat \mu(u,v)$ exhibits
various elegant modular and quasi-periodicity properties
\cite{Zwegers-thesis}. In particular, $\widehat \mu$ transforms under $SL(2,\IZ)$
transformations as a Jacobi form of weight one, and matrix-valued index
  $\frac{1}{2}\small{\left(\begin{array}{cc} -1 & 1 \\ 1 & -1  \end{array}\right)}$:
\be
\widehat \mu\!\left(\frac{a\tau+b}{c\tau+d},\frac{u}{c\tau+d},\frac{v}{c\tau+d}\right)
=\varepsilon(\gamma)^{-3}(c\tau+d)^{\frac{1}{2}}\, e^{-\pi i c(u-v)^2/(c\tau+d)}\widehat \mu(\tau,u,v), \non
\ee
where $\varepsilon(\gamma)$ corresponds to the multiplier system of the
$\eta$ function (\ref{eta}).\\

\noi {\bf Proof}\\
The proof of Equations (\ref{Acomp}) and (\ref{Rutau}) is originally given in Zwegers' thesis
\cite{Zwegers-thesis}. To aid the reader with the proof in Section \ref{AppellSign22}, we give here a proof based on the techniques of indefinite theta
functions discussed in Section \ref{indeftheta}. This technique will
also be used for the Appell-Lerch sums of signature $(2,2)$. We first expand the denominator as a geometric series, then
\be  
\label{Aexp}
A(u,v)=\half e^{\pi i u}\sum_{n\in \Zint, \ell\in \Zint}
(\sgn(\ell+\epsilon)+\sgn(n+{\rm Im}(u)/y))\,(-1)^n q^{n(n+1)/2+\ell
  n}e^{2\pi i \ell u +2\pi i n v},
\ee
where $0<\epsilon\ll 1$.  We see that the quadratic form $Q$ is given by
$Q(k)=n^2+2nl$ with $k=(n,\ell)$, and the bilinear form by $B(k,(u,v-u))=\ell u
+ nv$. Comparing with Equation (\ref{Theta}) and (\ref{Phi1}), we
furthermore derive that the vectors $C$ and $C'$ equal $C=(1,-1)$ and $C'=(0,-1)$. Note that in
Equation (\ref{Theta}), $b$ in the argument of $\Phi_1$ equals
$\Im(z)/y$. However in Equation (\ref{Aexp}), $\epsilon$ is not
related to $u$ and $v$. To express $A$ in terms of an indefinite theta
series (\ref{Theta}), we write Equation (\ref{Aexp}) as
\begin{eqnarray}
\label{Aexp2}
&&A(u,v)=\half \sum_{n\in \Zint, \ell\in \Zint+\half}
(\sgn(\ell+{\rm Im}(v-u)/y)+\sgn(n+{\rm Im}(u)/y))\,(-1)^n q^{n^2/2+\ell
  n}e^{2\pi i \ell u +2\pi i n v}\non \\
&&+\frac{1}{2}\sum_{n\in \mathbb{Z},\ell\in \Zint+\half}
\left(\sgn(\ell-\textstyle \half + \eps)-\sgn(\ell+{\rm Im}(v-u)/y)\right) (-1)^n q^{n^2/2+\ell
  n}e^{2\pi i \ell u +2\pi i n v},
\end{eqnarray}
where we shifted $\ell\mapsto \ell-\half$. Choosing $\epsilon=\half$,
and shifting $n\mapsto n-\ell$, brings the second line to the
form
\be
-\frac{i}{2}\theta_1(v)\,\sum_{\ell\in \Zint+\half}
\left(\sgn(\ell)-\sgn(\ell+{\rm Im}(v-u)/y)\right)  (-1)^{\ell-\half} q^{\ell^2/2}e^{2\pi i \ell (u-v)}.
\ee
Completing $A(u,v)$ now amounts to replacing the $\sgn(\dots)$'s on
the first line by $E_1(\dots)$'s, and by subtracting the second line in
Equation (\ref{Aexp2}), because this line is not modular. Since $Q(C',C')=0$, this
reproduces Equations (\ref{Acomp}) and (\ref{Rutau}) as
claimed. \hfill{ $\square$}\\

We will meet later three variations of the classical Appell function $A$
(\ref{Appell}) in the refined $U(3)$ VW partition function. See
Equations (\ref{g31-1}) and (\ref{f30}). In preparation for Section
\ref{VWP2}, we will determine here their completions. The three
functions $A_j(\tau,z)=A_j(z)$ are:
\be 
\label{Adef}
\begin{split}
A_0(z)&=-\tfrac{1}{2}\theta_3(6\tau,6z)+\sum_{k\in \IZ}
\frac{w^{-6k}q^{3k^2}}{1-w^6 q^{3k}}, \\
A_1(z)&= \sum_{k\in \IZ} \frac{w^{-6k+6}q^{3k^2-\frac{1}{3}}}{1-w^6q^{3k-1}},\\
A_2(z)&= \sum_{k\in \IZ} \frac{w^{-6k+6}q^{3k^2+3k+\frac{2}{3}}}{1-w^6q^{3k+1}},
\end{split}
\ee
with $w=e^{2\pi i z}$. \\

\noi {\bf Completion of $A_j$}\\
We will list the completions first and then
give the derivation based on the completion $\widehat \mu$
(\ref{Acomp}). We express the
completions $\widehat A_j$ of $A_j$ as\footnote{We will in
  the following suppress the variable $\tau$ from the arguments of the
non-holomorphic terms such as $R_{A_\alpha}(z)$} 
$$\widehat A_j(z)=A_j(z)-\frac{1}{2}R_{A_j}(z).$$
The non-holomorphic terms $R_{A_j}(z)$ equal
\be
\label{Aicomp}
\begin{split}
R_{A_0}(z)=&\theta_3(6\tau,6z)
R_{1,0}(6z)+\theta_2(6\tau,6z) R_{1,\frac{1}{2}}(6z),\\
R_{A_1}(z)=&\theta_3(6\tau,6z)
R_{1,\frac{1}{3}}(6z)+\theta_2(6\tau,6z) R_{1,\frac{5}{6}}(6z),\\
R_{A_2}(z)=&\theta_3(6\tau,6z)
R_{1,\frac{2}{3}}(6z)+\theta_2(6\tau,6z) R_{1,\frac{1}{6}}(6z),\\
\end{split}
\ee
where the $R_{1,\alpha}$ are defined as:
\be
\label{R1j}
R_{1,\alpha}(z):=\sum_{\ell\in \mathbb{Z}+\alpha}[\sign(\ell)-E_1(\sqrt{3y}(2\ell-a))]\,e^{6\pi i \ell z}q^{-3\ell^2}.
\ee
\\
\noi {\bf Proof}\\
Let us demonstrate this completion for $A_0$. We first rewrite $A_0$ as 
$$
A_0(z)=-\tfrac{1}{2}\theta_3(6z,6\tau)+\sum_{k\in \IZ}\left(\frac{w^{-6k}q^{3k^2}}{1-w^{12}q^{6k}}+\frac{w^{-6k+6}q^{3k^2+3k}}{1-w^{12}q^{6k}}\right),
$$
by multiplying the numerator and denominator in the sum by
$1+w^6q^{3k}$. Now we can express $A_0$ in terms of the original $A$ (\ref{Appell}):
\be
A_0(z)=-\tfrac{1}{2}\theta_3(6\tau,6z)+w^{-6}A(6\tau, 12z, -\tfrac{1}{2}-6z-3\tau)+A(6\tau, 12z, -\tfrac{1}{2}-6z)
\ee
Application of Equations (\ref{muuv}) and (\ref{Acomp}) gives then
the first line of Equation (\ref{Aicomp}). \hfill $\square$

\subsection{Example of signature $(2,1)$: the function $\Phi$}
We will consider in this subsection the function $\Phi$, which
is of signature $(2,1)$ and will occur in the refined
VW partition functions (\ref{f2jmu}). We define the
function $\Phi(\tau,u,v):=\Phi(u,v)$ by\footnote{We hope there will be no confusion between
this $\Phi$ and the kernel $\Phi_j$ used in the previous section.}
\be
\label{mu21}
\Phi(u,v):=\frac{1}{2}+\frac{e^{2\pi
  i u}}{b_{3,0}(\tau,v)}\sum_{k_1,k_2\in
  \Zint}\frac{e^{2\pi i
    v(k_1+2k_2)}q^{k_1^2+k_2^2+k_1k_2+2k_1+k_2}}{1-e^{2\pi i u}q^{2k_1+k_2}},
\ee
where $b_{3,0}$ is defined in Equation (\ref{b3j}). Note that $\Phi$ involves 
the same quadratic form in the numerator as is associated to
$b_{3,0}$, namely the quadratic form of the $\operatorname{A}_2$ root lattice.   

We will now show that this function can be expressed in terms of the
classical Appell-Lerch sum $\mu$ (\ref{muuv}), such that its completion and other properties can be
readily determined. To this end, we make the change of variables $k=k_2$,
$\ell=2k_1+k_2$ in the sum in Equation (\ref{mu21}), such that $\Phi$ takes the form
\be
\begin{split}
\Phi(u,v)&=\frac{1}{2}+\frac{e^{2\pi i u}}{b_{3,0}(\tau,v)}\left(\sum_{\ell,k \in 2\mathbb{Z}}+ \sum_{\ell,k \in
  2\mathbb{Z}+1} \right)\frac{e^{\pi i
  v(\ell+3k)}q^{\frac{1}{4}\ell^2+\ell+\frac{3}{4}k^2}}{1-e^{2\pi
  i u}q^\ell}
\\
&=\frac{1}{2}+\frac{e^{2\pi i u}}{b_{3,0}(\tau,v)}\left( \theta_3(6\tau,3v) \sum_{\ell\in \Zint}
\frac{e^{2\pi i v \ell}q^{\ell^2+2\ell}}{1-e^{2\pi i u}q^{2\ell}}+\theta_2(6\tau,3v)\sum_{\ell\in
\Zint}\frac{e^{2\pi i v (\ell-\frac{1}{2})}q^{\ell^2+\ell-\frac{3}{4}}}{1-e^{2\pi
i u}q^{2\ell-1}}\right).
\end{split}
\ee
The sums over $\ell$ on the second line can now be replaced by
specializations of $\mu$. We arrive in this way at
\be
\label{Phi_mu}
\begin{split}
\Phi(u,v)=\frac{1}{2}\,+\,&\frac{e^{\pi i u}}{b_{3,0}(\tau,v)} \left\{\textstyle
\theta_3(6\tau,3v)\,\theta_1(2\tau,v+\frac{1}{2}+\tau)\,\mu(2\tau,u,v+\frac{1}{2}+\tau)\right.\\
&\left.+e^{-\pi i
v}q^{-\frac{1}{4}}\,\theta_2(6\tau,3v)\,\theta_1(2\tau,v+\tfrac{1}{2})\,\mu(2\tau,u-\tau,v+\tfrac{1}{2})\right\}.
\end{split}
\ee
We will simplify Equation (\ref{Phi_mu}) further for later reference. Using the
following identity for $\mu$ \cite{Zwegers-thesis},
\be 
\label{appellperiod}
\mu(u+z,v+z) - \mu(u,v)
=-\frac{i\eta^3\,\theta_1(u+v+z)\,\theta_1(z)}
{\theta_1(u)\,\theta_1(v)\,\theta_1(u+z)\,\theta_1(v+z)},
\ee
we can express $\mu(2\tau, u-\tau,v+\tfrac{1}{2})$
in terms of $\mu(2\tau, u,v+\frac{1}{2}+\tau)$. Upon using also the
relation (\ref{cubictheta}) for $b_{3,0}$, $\Phi$ can be expressed as
\be
\label{Phi_mu2}
\begin{split}
\Phi(u,v)=&\frac{1}{2}-e^{\pi i (u-v)}
q^{-\frac{1}{4}}\mu(2\tau, u-\tau,v +{\textstyle \frac{1}{2}})\\
&-i \frac{\,\eta(2\tau)^3\,\theta_4(2\tau,0)\,
\theta_3(6\tau,3v)\,\theta_2(2\tau,u+v)}{\theta_1(2\tau,u)\,\theta_4(2\tau, u)\,\theta_2(2\tau, v)\,b_{3,0}(\tau, v)}.
\end{split}
\ee

For the refined VW partition functions, we will later be interested in the specializations
$\Phi(4z,-2z)$ and $\Phi(4z,-2z-\tau)$. Using Equation (\ref{Phi_mu2}), these
can be expressed as:
\be
\label{mu211}
\Phi(4z,-2z)=\frac{1}{2}-w^3q^{-\frac{1}{4}}\mu(2\tau, 4z-\tau,-2z+\textstyle
\frac{1}{2})-\frac{i \,\eta(\tau)^3\,\theta_3(6\tau, 6z)}{\theta_1(\tau, 4z)\,b_{3,0}(\tau,2z)},
\ee
and
\be
\label{mu212}
\Phi(4z,-2z-\tau)=\frac{1}{2}-w^3q^{\frac{1}{4}}\mu(2\tau, 4z-\tau,-2z-\tau+\textstyle\frac{1}{2})
-\frac{i w^3q^{\frac{1}{4}}\eta(\tau)^3\,\theta_2(6\tau,6z)}{\theta_1(\tau,4z)\,b_{3,0}(\tau,2z)}.
\ee
The completions of $\Phi$ follow immediately by replacing $\mu$ by
$\widehat \mu$.

\subsection{Examples of signature $(2,2)$: the functions $\Psi_0$ and $\Psi_1$}
\label{AppellSign22}
We consider next two Appell-Lerch sums of signature $(2,2)$. The
positive definite lattice corresponds again to the ${\rm A}_2$ root
lattice as before. The two functions $\Psi_{j}: \mathbb{H}\times \mathbb{C}^2\to
\mathbb{C}$ are defined by 
\be
\label{Psi01} 
\begin{split}
\Psi_0(\tau,u,v)&:=\frac{1}{4}+\frac{e^{2\pi i
    u}}{b_{3,0}(\tau,v)}\sum_{k_1,k_2\in \mathbb{Z}}\frac{e^{2\pi i
    v(k_1+2k_2)}q^{k_1^2+k_2^2+k_1k_2+2k_1+k_2}}{(1-e^{2\pi i
    u}q^{2k_1+k_2})(1-e^{2\pi i u}q^{k_2-k_1})}, \\
\Psi_1(\tau,u,v)&:=\frac{e^{2\pi i
    u}}{b_{3,0}(\tau,v)}\sum_{k_1,k_2\in \mathbb{Z}-\frac{1}{3}}\frac{e^{2\pi i
    v(k_1+2k_2)}q^{k_1^2+k_2^2+k_1k_2+k_1+k_2}}{(1-e^{2\pi i
    u}q^{2k_1+k_2})(1-e^{2\pi i u}q^{k_2-k_1})}.
\end{split}
\ee
For simplicity, we have included only two elliptic variables $u$ and $v$ in
$\Psi_j$. One could straightforwardly refine the function by including
two more elliptic variables as in Equation (\ref{AQ}).\\
 
\noi {\bf Completion of $\widehat \Psi_j$}\\
To state the modular completions $\widehat \Psi_j$ of $\Psi_j$, we define $R_{2,\mu}(z)$ by
\be
\label{compR}
\begin{split}
R_{2,\mu}(z)&:=\sum_{k_3,k_4\in \mathbb{Z}^2+\mu}[ \sgn(k_3)\,\sgn(k_4)-(\sgn(k_4) -E_1(\sqrt{3y}(k_4-a)))\,\sgn(2k_3-k_4)
\\
&-(\sgn(k_3)
-E_1(\sqrt{3y}(k_3-a)))\,\sgn(2k_4-k_3)\\ 
&-E_2\!\left(\tfrac{1}{\sqrt{3}};\sqrt{y}
  (2k_3-k_4-a), \sqrt{3y} (k_4-a)\right)]\\
& \times  e^{2\pi i(k_3+k_4)z}q^{-k_3^2-k_4^2+k_3k_4}
\end{split}
\ee
with $\mu\in \IR^2$ and $a=\Im(z)/y$. The completions $\widehat
\Psi_j$, $j=0,1$, are then defined by 
\be
\begin{split}
&\widehat \Psi_j:=\Psi_j-\textstyle{\frac{1}{4}} R_{\Psi_j},
\end{split}
\ee
with
\be
\label{RPsi0}
\begin{split}
R_{\Psi_0}(u,v)&:=R_{2,0}(u-v)  \\
& +4 \Phi(u,v)\,R_{1,0}(u-v)
+4\,e^{\pi i (v-u)}q^{-\frac{1}{4}}\left(
  \Phi(u+\tau,v)-\tfrac{1}{2}\right)\, R_{1,\frac{1}{2}}(u-v),
\end{split}
\ee
and
\be
\label{RPsi1}
\begin{split}
R_{\Psi_1}(u,v)=&R_{2,\frac{1}{3}(-1,1)}(u-v)\\
& +2\, \Phi(u,v)\left( R_{1,\frac{1}{3}}(u-v)+R_{1,\frac{2}{3}}(u-v)\right) \\
&+2\,e^{\pi i (v-u)}q^{-\frac{1}{4}}\left(
  \Phi(u+\tau,v)-\tfrac{1}{2}\right)\left( R_{1,\frac{1}{6}}(u-v)+R_{1,\frac{5}{6}}(u-v)\right).
\end{split}
\ee
Then the $\widehat \Psi_j$ transform as a two-variable Jacobi form of weight 1 under the modular group
$\Gamma(3)$ and index $m=-\left(\begin{array}{cc} 1 & 1\\ 1 &
    1 \end{array}\right)$. The completion of $\Psi_0$ was earlier
proved in \cite[Theorem 5.3]{Alexandrov:2016enp}. In the following, we
give a proof for the completion of $\Psi_1$.\\

\noi {\bf Proof}\\
The proof follows the strategy of the proof of the completion of
$\mu$ in Section \ref{muA}, and is very similar to the proof for the
completion of $\Psi_0$ in Reference
\cite{Alexandrov:2016enp}. We first write the denominators in the
summand of $\Psi_1$ as a geometric sum. In this way, we can relate $\Psi_1$ to an
indefinite theta series whose associated lattice has signature
(2,2). Using the techniques of References \cite{Alexandrov:2016enp, Vigneras:1977}, we then determine the modular
completion of the indefinite theta series, and consequently of $\Psi_1$.

Let us start by considering the Appell function $C(\tau,u,v):=C(u,v)$ obtained from
$\Psi_1$ by multiplying with $b_{3,0}$,
\be
\begin{split}
C(u,v)&=e^{2\pi i u}\sum_{k_1,k_2\in \IZ-\frac{1}{3}}
\frac{e^{2\pi i v(k_1+2k_2)}q^{k_1^2+k_2^2+k_1k_2+k_1+k_2}}{(1-e^{2\pi
  i u} q^{2k_1+k_2})(1-e^{2\pi i u}q^{k_2-k_1})}
\end{split}
\ee
We expand the denominator of $C$ using a double geometric series, 
\be
\begin{split}
C(u,v)=&\tfrac{1}{4}\sum_{k_1,k_2\in \IZ-\frac{1}{3} \atop k_3,k_4\in
  \IZ}
[\sgn(k_3+\epsilon)+\sgn(2k_1+k_2+a)][\sgn(k_4+\epsilon)+\sgn(k_2-k_1+a)]\\
& \times e^{2\pi i v(k_1+2k_2)+u(k_3+k_4+1)}q^{k_1^2+k_2^2+k_1k_2+(2k_1+k_2)(k_3+\frac{2}{3})+(k_2-k_1)(k_4+\frac{1}{3})},
\end{split}
\ee
where $a=\mathrm{Im}(u)/y$ and $0<\eps<1$. After shifting $k_3\mapsto k_3-\frac{2}{3}$ and $k_4\mapsto
k_4-\frac{1}{3}$, $C$ takes the form
\be
\label{Cuv}
\begin{split}
C(u,v)=&\tfrac{1}{4}\sum_{k\in \IZ^4+\nu}
[\sgn(k_3-\tfrac{2}{3}+\epsilon)+\sgn(2k_1+k_2+a)][\sgn(k_4-\tfrac{1}{3}+\epsilon)+\sgn(k_2-k_1+a)]\\
& \times e^{2\pi i v(k_1+2k_2)+2\pi i u(k_3+k_4)}q^{k_1^2+k_2^2+k_1k_2+(2k_1+k_2)k_3+(k_2-k_1)k_4)},
\end{split}
\ee
with $\nu=\frac{1}{3}(-1,-1,-1,1)$. We can express the second line
more compactly by introducing the bilinear form $B(x,y)$ with matrix representation
\be
{\rm B}=\left(\begin{array}{cccc} 2 & 1 & 2 & -1 \\ 1 & 2 & 1 & 1\\ 2 & 1 & 0 & 0 \\ -1 & 1 & 0 & 0 \end{array} \right)\ ,
\ee
and associated quadratic form $Q(k)=B(k,k)$. The second line can then
be expressed as $e^{2\pi i B(k,z)}q^{Q(k)/2}$ with $z=(0,u,v-u,v-u)\in \IC^4$. The
right hand side of Equation (\ref{Cuv}) is almost of the form of an
indefinite theta series of signature (2,2). However, the argument of the
$\sgn(\cdots)$'s in Equation (\ref{Cuv}) should then match with the shift
$\Im(z)/y$ as in Equation (\ref{Theta}). This is not
the case for the $\sgn(\cdots)$'s with $k_3$ and $k_4$. Nevertheless, we can
write $C(u,v)$ as an indefinite theta 
function plus a correction term:
\be
\label{Cuvexp}
\begin{split}
C(u,v)=&\tfrac{1}{4}\sum_{k\in \IZ^4+\mu}
[\sgn(k_3+b-a)+\sgn(2k_1+k_2+a)][\sgn(k_4+b-a)+\sgn(k_2-k_1+a)]\\
&\times e^{2\pi i B(k,z)}q^{Q(k)/2}+ s(k,z,\eps)\,e^{2\pi i B(k,z)}q^{Q(k)/2},
\end{split}
\ee
where $b=\Im(v)$ and $s(k,z,\eps)$ is given by 
\be
\begin{split}
s(k,z,\eps)=&[\sgn(k_3-\tfrac{2}{3}+\eps)-\sgn(k_3+b-a)]\,\sgn(k_2-k_1+a)\\
&+[\sgn(k_4-\tfrac{1}{3}+\eps)-\sgn(k_4+b-a)]\,\sgn(2k_1+k_2+a)\\
&+\sgn(k_3-\tfrac{2}{3}+\eps)\sgn(k_4-\tfrac{1}{3}+\eps)-\sgn(k_3+b-a)\sgn(k_4+b-a).
\end{split}
\ee
Note that in $s(k,z,\eps)$, we may replace $-\frac{2}{3}+\eps$
(respectively $-\frac{1}{3}+\eps$) by 0, since this does not change
the value of $\sgn(k_3-\frac{2}{3}+\eps)$ for any $k_3\in \IZ-\frac{1}{3}$
(respectively the value of $\sgn(k_4-\frac{1}{3}+\eps)$ for any $k_4\in \IZ+\frac{1}{3}$).

We write the modular completion $\widehat C$ of $C$ as
\be
\widehat C=C-\tfrac{1}{4} R_C,
\ee  
where $R_C$ is a subleading non-holomorphic function. To determine
$R_C$, we complete the first line of Equation
(\ref{Cuvexp}) using the techniques of indefinite theta series
\cite{Alexandrov:2016enp}, and moreover subtract the non-modular
second line of (\ref{Cuvexp}). In this way, one derives for $R_C$:
\be
\label{RCuv}
\begin{split}
R_C(u,v)=&\sum_{k\in \IZ+\mu} \left\{ \left(\sgn(k_3)-E_1(\sqrt{3y}(k_3+b-a) \right)\,\sgn(k_2-k_1+a)) \right.\\
&  +\left(\sgn(k_4)-E_1(\sqrt{3y}(k_4+b-a) \right)\,\sgn(2k_1+k_2+a)) \\
& \left.
  +\,\sgn(k_3)\,\sgn(k_4)-E_2\!\left(\tfrac{1}{\sqrt{3}};\sqrt{y}
    (2k_3-k_4+b-a),\sqrt{3y}(k_4+b-a) \right)     \right\}\\
& \times e^{2\pi i B(k,z)}q^{Q(k)/2}.
\end{split}
\ee

Next we want to carry out the geometric sums in this equation. To this
end, we combine the first and second line of (\ref{RCuv}), using the
transformation
\be
k_1\mapsto -k_1,\quad k_2\mapsto k_2+k_1,\quad k_3\mapsto k_4,\quad
k_4\mapsto k_3,
\ee 
which leaves both $B(k,z)$ and $Q(k)$ invariant. Note that this
transformation does flip the conjugacy class $\mu$ to $-\mu$. Thus we
can write the first two lines of $R_C$ as
\be
\label{RCfirsttwo}
\sum_{k\in \IZ\pm\mu} \left(\sgn(k_4)-E_1(\sqrt{3y}(k_4+b-a) \right)\,\sgn(2k_1+k_2+a) \,e^{2\pi i B(k,z)}q^{Q(k)/2}.
\ee
Our next aim is to express this sum in terms of $R_{1,\alpha}$ and to
carry out the sum over $k_3$ as a geometric series. The sum over $k_3$
will in fact return the specializations of $\Phi(u,v)$ we met earlier
in Equation (\ref{mu21}). We start with replacing in Equation (\ref{RCfirsttwo}) $\sgn(2k_1+k_2+a)$ by
$\sgn(2k_1+k_2+a)+\sgn(2k_3-k_4)-\sgn(2k_3-k_4)$. We can then carry out the
sum over $k_3$ in the terms multiplying
$\sgn(2k_1+k_2+a)+\sgn(2k_3-k_4)$,
\be
\label{partresum}
\sum_{k_1,k_2,k_3\in \IZ\mp\frac{1}{3}}
(\sgn(2k_1+k_2+a)+\sgn(2k_3-k_4)) e^{2\pi i B(k,z)}q^{Q(k)/2}.
\ee
To this end, we shift the summation variables as follows:
\be
k_1\mapsto k_1+\tfrac{1}{2} k_4,\quad k_2\mapsto k_2-k_4,\quad k_3
\mapsto k_3,\quad k_4\mapsto k_4,
\ee
such that Equation (\ref{partresum}) becomes:
\be
\begin{split}
\sum_{k_2\in \IZ, k_3 \in \IZ\mp\frac{1}{3}\atop k_1+\frac{1}{2}k_4 \in
  \IZ\mp\frac{1}{3}} & (\sgn(2k_1+k_2+a)+\sgn(2k_3-k_4)) \\
&\times e^{2\pi i v(k_1+2k_2-\frac{3}{2}k_4)+2\pi i u (k_3+k_4)}q^{k_1^2+k_2^2+k_1k_2+(2k_1+k_2)(k_3-\frac{1}{2}k_4)-\frac{3}{4}k_4^2}.
\end{split}
\ee
There are two possibilities for $k_4$, $k_4\in 2\IZ\pm\frac{1}{3}$ and
$k_4\in 2\IZ\pm\frac{4}{3}$. Let us first assume that $k_4\in
2\IZ\pm\frac{1}{3}$. Then, since $k_3\in
\IZ\mp\frac{1}{3}$, $\sgn(2k_3-k_4)$ is positive if $k_3\geq
\frac{k_4}{2}+\frac{1}{2}$ and negative for $k_3<\frac{k_4}{2}+\frac{1}{2}$. After resumming
$k_3$, we arrive therefore at:
\be
\label{RCtwolinesk41}
\begin{split}
&2\sum_{k_1\in \IZ+\frac{1}{2} \atop k_2\in \IZ} \frac{e^{2\pi i
  v(k_1+2k_2-\frac{3}{2}k_4)+2\pi i u
  (\frac{3k_4}{2}+\frac{1}{2})}q^{k_1^2+k_2^2+k_1k_2+\frac{1}{2}(2k_1+k_2)-\frac{3}{4}k_4^2}}{1-e^{2\pi i
u} q^{2k_1+k_2}}\\
&=2e^{\pi i (v-u)+2\pi
  i\frac{3}{2}k_4(u-v)}q^{-\frac{1}{4}-\frac{3}{4}k_4^2}b_{3,0}(v) (\Phi(u+\tau,v)-\tfrac{1}{2}),
\end{split}
\ee 
where we substituted $\Phi$ (\ref{mu21}). The second possibility is $k_4\in 2\IZ\pm\frac{4}{3}$. Carrying out the sum
over $k_3\geq \frac{k_4}{2}$ and $k_3<\frac{k_4}{2}$ gives in this case:
\be
\label{RCtwolinesk42}
\begin{split}
&2\sum_{k_1,k_2\in \IZ} \frac{e^{2\pi i v(k_1+2k_2)+2\pi
  i \frac{3}{2}k_4(u-v)}q^{k_1^2+k_2^2+k_1k_2-\frac{3}{4}k_4^2}}{1-e^{2\pi i u}q^{2k_1+k_2}}\\
&-\sum_{k_1,k_2\in \IZ} e^{2\pi i v(k_1+2k_2)+2\pi
  i \frac{3}{2}k_4(u-v)}q^{k_1^2+k_2^2+k_1k_2-\frac{3}{4}k_4^2},
\end{split}
\ee
which equals
\be
2b_{3,0}(v) \Phi(u,v) e^{2\pi i (u-v)\frac{3k_4}{2}} q^{-\frac{3}{4}k_4^2}.
\ee

As a result, we can express the sum over $k_4$ in the contributions of
these terms to Equation (\ref{RCfirsttwo}) in terms of
$R_{1,\alpha}$. Combining all terms we find that the first two lines
of Equation (\ref{RCuv}) equals
\be
\label{2ndline}
\begin{split}
& 2
b_{3,0}(v)\,(R_{1,\frac{2}{3}}(u-v)+R_{1,\frac{1}{3}}(u-v))\,\Phi(u,v)\\
& +2 e^{\pi i
  (v-u)}q^{-\frac{1}{4}}\, b_{3,0}(v)\, (R_{1,\frac{1}{6}}(u-v)+R_{1,\frac{5}{6}}(u-v))\,
(\Phi(u+\tau,v)-\tfrac{1}{2}) \\
&-  \sum_{k\in \IZ^2\pm\mu} \left(\sgn(k_4)-E_1(\sqrt{3y}(k_4+b-a)
\right)\sgn(2k_3-k_4) e^{2\pi i B(k,z)} q^{Q(k)/2}. 
\end{split}
\ee
Using the transformations:
\be
k_1\mapsto k_1-k_3+k_4,\quad k_2\mapsto k_2,\quad k_3\mapsto k_3,
\quad k_4\mapsto k_4,
\ee
the sum over $k_{1,2}$ decouples from the sum over $k_{3,4}$, and can
be factored out as $b_{3,0}$. The third line of Equation (\ref{RCuv}) and the
third line of Equation (\ref{2ndline}) combine to $b_{3,0}\,R_{2,\frac{1}{3}(-1,1)}$
(\ref{compR}). Adding up all contributions and dividing by $b_{3,0}$ gives
the desired result. \hfill $\square$

\subsection{Taylor expansions and iterated period integrals}
\label{Taylerexp}
A building block for the refined VW partition function $f_{3,\mu}$ is the 
specialization $\Psi_\mu(4z,-2z)$. The unrefined partition
function involves the second Taylor coefficient in $z$ of these
functions. To derive the completion, we determine in this subsection
the second derivative of the completion $R_{\Psi_j}$ (\ref{compR}) at $z=0$. 
To this end, we need to determine the first derivative of
$R_{1,\alpha}$ and the second derivative of $R_{2,\alpha}$. We will
express these derivatives as (iterated) integrals of theta series.\\

\noi {\bf Taylor expansion of $R_{1,\alpha}$}\\
We start with $R_{1,\alpha}$ (\ref{R1j}) and write its Taylor expansion as
\be
R_{1,\alpha}(z)=\sum_{\ell\geq 0} R^{(\ell)}_{1,\alpha}\,\frac{z^\ell}{\ell!}.
\ee
The first two coefficients are given by:
\be
\label{R1alphader}
\begin{split}
R^{(0)}_{1,\alpha}&=-\sum_{\ell\in \mathbb{Z}+\alpha}M_1(2\sqrt{3y}\ell)\,q^{-3\ell^2},\\
R^{(1)}_{1,\alpha}&=-\sqrt{\frac{3}{2}} \int_{-\bar \tau}^{i\infty} \frac{\sum_{\ell\in
  \mathbb{Z}+\alpha} e^{6\pi i
    \ell^2 w}}{(-i(w+\tau))^{\frac{3}{2}}}dw.
\end{split}
\ee
\\
\noi {\bf Proof}\\
The expression for $R_{1,\alpha}^{(0)}$ follows immediately from the
definition (\ref{R1j}) and Equation (\ref{E1M}). Note that for $\alpha=0$ and
$\alpha=\frac{1}{2}$, $R_{1,\alpha}$ vanishes. 

To determine the first derivative, $R_{1,\alpha}^{(1)}$, we first set
$F(z)=E_1(\sqrt{3y}(2\ell-a))\,e^{6\pi i \ell z}$. For its first
derivative, $F^{(1)}(0)$, we find:
\be
F^{(1)}(0)=i\sqrt{\frac{3}{y}}\,e^{-12\pi y \ell^2}+6\pi i \ell\, E_1(2\sqrt{3y}\ell).
\ee
Substitution in $R^{(1)}_\alpha$ gives:
\be
R^{(1)}_\alpha=-\sum_{\ell\in \IZ+\alpha} \left[\sqrt{\tfrac{3}{y}} \,
 e^{-12\pi y \ell^2}+6\pi i \ell\, M_1(2\sqrt{3y}\ell) \right] q^{-3\ell^2}.
\ee
Finally, substitution of Equation (\ref{uM1}) gives the expression 
of Equation (\ref{R1alphader}). \hfill $\square$
\\ 

\noi {\bf Taylor expansion of $R_{2,\beta}$}\\
We consider next the Taylor expansion of $R_{2,\beta}(z)$ around
$z=0$: $$R_{2,\beta}(z)=\sum_{\ell\geq 0}
R^{(\ell)}_{2,\beta}\,\frac{z^\ell}{\ell!}.$$
Our interest is in two choices of $\beta$: $\beta=0$ and
$\frac{1}{3}(-1,1)$. Since $R_{2,\beta}(z)$ is a symmetric
function in $z$ for both choices, only
even derivatives of $R_{2,\beta}$ are non-vanishing at $z=0$. We have \\
\be
\label{R208}
\begin{split}
R^{(0)}_{2,\beta}&=\left\{ \begin{array}{rl}-\frac{1}{3}, & \qquad \beta=0,\\
    0, & \qquad  \beta=\frac{1}{3}(-1,1), \end{array}\right.\\
R^{(1)}_{2,\beta}&=0,\\
R^{(2)}_{2,\beta}&=-2\sqrt{3}\int_{-\bar \tau}^{i\infty} \int_{w_2}^{i\infty}
\frac{\sum_{k_3,k_4\in \IZ^2+\beta} e^{\frac{\pi
      i}{2}(2k_3-k_4)^2w_1+\frac{3\pi
      i}{2}k_4^2w_2}}{\sqrt{-(w_1+\tau)^3(w_2+\tau)^3}}dw_1dw_2.
\end{split}
\ee

\noi {\bf Proof}\\
It follows
from Equations (\ref{E2M}) and (\ref{E1M}), that the constant term equals
\be
R^{(0)}_{2,\beta}=-\sum_{k_3,k_4\in \IZ^2+\beta}
M_2\!\left(\tfrac{1}{\sqrt{3}};\sqrt{y} (2k_3-k_4), \sqrt{3y}
  k_4\right) q^{-k_3^2-k_4^2+k_3k_4}.
\ee
This can be further simplified. To this end, we express the summand as
an iterated period integral. For $2k_3-k_4\neq 0$ and
$2k_4-k_3\neq 0$, this gives using Equation (\ref{E2period}):
\be  
\begin{split}
&M_2\!\left(\tfrac{1}{\sqrt{3}};\sqrt{y} (2k_3-k_4),
  \sqrt{3y}  k_4\right) q^{-k_3^2-k_4^2+k_3k_4}=\\
&-\tfrac{\sqrt{3}}{2} (2k_3-k_4)k_4 \int_{-\bar
  \tau}^{i\infty}\int_{w_2}^{i\infty}  \frac{e^{\frac{\pi i}{2}(2k_3-k_4)^2
    w_1+\frac{3\pi i}{2}k_4^2w_2}}{\sqrt{-(w_1+\tau)(w_2+\tau)}} dw_1 dw_2\\
&-\tfrac{\sqrt{3}}{2} (2k_4-k_3)k_3 \int_{-\bar
  \tau}^{i\infty}\int_{w_2}^{i\infty}  \frac{e^{\frac{\pi i}{2}(2k_4-k_3)^2 
    w_1+\frac{3\pi i}{2} k_3^2 w_2}}{\sqrt{-(w_1+\tau)(w_2+\tau)}} dw_1 dw_2.
\end{split}
\ee
Since we sum over all $k_3,k_4\in \IZ^2+\beta$, we can change in the
third line $k_3 \to -k_4$ and $k_4\to k_3-k_4$. Then the third line is
cancelled by the second line. One proves similarly that the only
non-vanishing contribution is due to $k_3=k_4=0$. From Equation
(\ref{u1u20}), we deduce the desired result. 

To determine the second derivative, consider first
$K(z)=E_1(\sqrt{3y}(k_4-a))e^{2\pi i (k_3+k_4)z}$. We find for
the first derivative:
\be
\begin{split}
K^{(1)}(z)=&i\sqrt{\tfrac{3}{y}}\,e^{-3\pi yk_4^2-3\pi
  ya^2+2\pi i z (k_3-\frac{1}{2}k_4)+3\pi i \bar z k_4}\\
&+2\pi i (k_3+k_4)\,E_1(\sqrt{3y}(k_4-a))\, e^{2\pi i z (k_3+k_4)}.
\end{split}
\ee
Taking the second derivative at $z=0$, we arrive at
\be
\begin{split}
K^{(2)}=&-2\pi \sqrt{\tfrac{3}{y}}(2k_3+\tfrac{1}{2}k_4)e^{-3\pi
  yk_4^2} -4\pi^2(k_3+k_4)^2E_1(\sqrt{3y}k_4). 
\end{split}
\ee
Next we define: 
\be
L(z)=E_2\!\left(\tfrac{1}{\sqrt{3}};\sqrt{y} (2k_3-k_4-a), \sqrt{3y} (k_4-a)\right)\,  e^{2\pi\I(k_3+k_4)z}.
\ee
We find for its first derivative:
\be
\begin{split}
L^{(1)}(z)=&2\pi i (k_3+k_4)L(z)\\
&+i\sqrt{\tfrac{3}{y}}E_1(\sqrt{y}(2k_4-k_3-a))\,e^{-3\pi
  k_3^2y-3\pi y a^2+2\pi i (k_4-\frac{1}{2}k_3)z+3\pi i k_3\bar
  z}\\
&+i\sqrt{\tfrac{3}{y}}E_1(\sqrt{y}(2k_3-k_4-a))\,e^{-3\pi
  k_4^2y-3\pi y a^2+2\pi i (k_3-\frac{1}{2}k_4)z+3\pi i k_4\bar
  z}.
\end{split}
\ee
After differentiating one more time and setting $z=0$, we find for $L^{(2)}$:
\be
\begin{split}
L^{(2)}=&-4\pi^2 (k_3+k_4)^2 E_2\!\left(\tfrac{1}{\sqrt{3}};\sqrt{y}
  (2k_3-k_4), \sqrt{3y} k_4\right)\\
&-2\pi \sqrt{\tfrac{3}{y}}
(2k_4+\tfrac{1}{2}k_3)\,E_1(\sqrt{y}(2k_4-k_3))\,e^{-3\pi y k_3^2}\\
&-2\pi \sqrt{\tfrac{3}{y}}
(2k_3+\tfrac{1}{2}k_4)\,E_1(\sqrt{y}(2k_3-k_4))\,e^{-3\pi y k_4^2}\\
&-\frac{2\sqrt{3}}{y}e^{-4\pi y(k_3^2-k_3k_4+k_4^2)}.
\end{split}
\ee
After substitution of $K^{(2)}$ and $L^{(2)}$ in $R^{(2)}_{2,\mu}$, we arrive at: 
\be
\label{R20}
\begin{split}
R^{(2)}_{2,\beta}=&4\pi^2\sum_{k_3,k_4} (k_3+k_4)^2
M_2\!\left(\tfrac{1}{\sqrt{3}};\sqrt{y} (2k_3-k_4),
  \sqrt{3y}  k_4\right) q^{-k_3^2-k_4^2+k_3k_4}\\
&-4\pi \sqrt{\frac{3}{y}}\sum_{k_3,k_4}(2k_3+\tfrac{1}{2}k_4) \sgn(2k_3-k_4)q^{-\frac{1}{4}(2k_3-k_4)^2}\bar
q^{\frac{3}{4}k_4^2}\\
&+4\pi \sqrt{\tfrac{3}{y}}\sum_{k_3,k_4}(2k_3+\tfrac{1}{2}k_4)E_1(\sqrt{y}(2k_3-k_4))q^{-\frac{1}{4}(2k_3-k_4)^2}\bar
q^{\frac{3}{4}k_4^2}\\
&+\frac{2\sqrt{3}}{y}\,\bar b_{3,j_\beta},
\end{split} 
\ee  
where $j_\beta=0$ (respectively $j_\beta=1$) for $\beta=0$ (respectively
$\beta=\frac{1}{3}(1,-1)$). The first line in Equation (\ref{R20}) corresponds to combining the second term of
$K^{(2)}(0)$ and the first line of $L^{(2)}(0)$. The second line of Equation (\ref{R20})
is due to the first term of $K^{(2)}(0)$. Replacing the
$\sgn(\cdots)$ on the second line and $E_1$  on the third by $M_1$, we arrive at:
\be
\label{R205}
\begin{split}
R^{(2)}_{2,\beta}=&4\pi^2\sum_{k_3,k_4} (k_3+k_4)^2
M_2\!\left(\tfrac{1}{\sqrt{3}};\sqrt{y} (2k_3-k_4),
  \sqrt{3y}  k_4\right) q^{-k_3^2-k_4^2+k_3k_4}\\
&+4\pi \sqrt{\tfrac{3}{y}}\sum_{k_3,k_4}(2k_3+\tfrac{1}{2}k_4)M_1(\sqrt{y}(2k_3-k_4))q^{-\frac{1}{4}(2k_3-k_4)^2}\bar
q^{\frac{3}{4}k_4^2}\\ 
&+\frac{2\sqrt{3}}{y}\,\bar b_{3,j_\beta}.
\end{split}
\ee

In the following, we will write $R^{(2)}_{2,\beta}$ more concisely as
a single iterated period integral. To this end, we substitute the expression for $M_2$ as a period integral
(\ref{E2period}). Then the first line of Equation (\ref{R20}) becomes
\be
\label{R201}
-4\sqrt{3}\pi^2 \sum_{k_3,k_4} (k_3+k_4)^2 (2k_3-k_4)k_4 \int_{-\bar
  \tau}^{i\infty}\int_{w_2}^{i\infty}  \frac{e^{\frac{\pi i}{2}(2k_3-k_4)^2
    w_1+\frac{3\pi i}{2}k_4^2w_2}}{\sqrt{-(w_1+\tau)(w_2+\tau)}} dw_1 dw_2,
\ee
for $2k_3-k_4\neq 0$ and $0$ otherwise. To bring $R_{2,\beta}^{(2)}$ in a simpler form, we write $\sum_{k_3,k_4}$ in Equation (\ref{R201}) as $\tfrac{1}{2}\sum_{k_3,k_4}+\tfrac{1}{2}\sum_{k_3,k_4}$ and make the transformation $k_3\to k_4-k_3$ and $k_4\to k_4$
in the second sum. This shows that the first line of Equation (\ref{R20})
equals
\be
\label{R204}
-6\sqrt{3}\pi^2 \sum_{k_3,k_4} (2k_3-k_4)^2k_4^2 \int_{-\bar
  \tau}^{i\infty}\int_{w_2}^{i\infty}  \frac{e^{\frac{\pi i}{2}(2k_3-k_4)^2
    w_1+\frac{3\pi i}{2}k_4^2w_2}}{\sqrt{-(w_1+\tau)(w_2+\tau)}} dw_1 dw_2,
\ee
if $2k_3-k_4\neq 0$ and vanishes otherwise. To combine this line with
the other lines, let us partially integrate with respect to
$w_2$. This expresses Equation (\ref{R204}) as the sum of three integrals:
\be
\label{R206}
\begin{split}
&i4\sqrt{3}\pi\sum_{k_3,k_4} (2k_3-k_4)^2\\
&\left\{ - \frac{1}{\sqrt{2y}} \bar q^{\frac{3}{4}k_4^2}\int_{-\bar \tau}^{i\infty}
  \frac{e^{\frac{\pi
        i}{2}(2k_3-k_4)^2w_1}}{\sqrt{-i(w_1+\tau)}}dw_1\right.\\
&+\int_{-\bar \tau}^{i\infty} \frac{e^{\frac{\pi
      i}{2}(2k_3-k_4)^2w_2+\frac{3\pi i}{2}k_4^2w_2 }}{(-i(w_2+\tau))}dw_2\\
&\left.-\frac{i}{2}\int_{-\bar
  \tau}^{i\infty}\int_{w_2}^{i\infty}\frac{e^{\frac{\pi
      i}{2}(2k_3-k_4)^2w_1+\frac{3\pi i}{2}k_4^2w_2}}{\sqrt{(w_1+\tau)(w_2+\tau)^3}}dw_1dw_2\right\}.
\end{split}
\ee 

To partially integrate the second integral to $w_2$, note that, due the symmetry $k_3\leftrightarrow k_4$,  the factor
$(2k_3-k_4)^2$ can be replaced by $2(k_3^2+k_4^2-k_3k_4)$ for this
term. After this substitution, the second integral in Equation (\ref{R206}) can easily be partially
integrated to obtain:
\be
\begin{split}
&i8\sqrt{3}\pi\sum_{k_3,k_4} (k_3^2+k_4^2-k_3k_4) \int_{-\bar
  \tau}^{i\infty} \frac{e^{2\pi i (k_3^2+k_4^2-k_3k_4)w_2}}{(-i(w_2+\tau))}dw_2\\
&=-\frac{2\sqrt{3}}{y}\bar b_{3,j_\mu}-4i\sqrt{3}\sum_{k_3,k_4}\int_{-\bar
  \tau}^{i\infty}\frac{e^{2\pi i (k_3^2+k_4^2-k_3k_4)w_2}}{(-i(w_2+\tau))^2}dw_2.
\end{split} 
\ee
Next, we partially integrate the third integral of Equation (\ref{R206})
to $w_1$, which gives:
\be
\begin{split}
&i4\sqrt{3}\int_{-\bar\tau}^{i\infty} \frac{e^{2\pi i
    (k_3^2+k_4^2-k_3k_4)w_2}}{(-i(w_2+\tau))^2}dw_2\\
&-2\sqrt{3}\int_{-\bar \tau}^{i\infty} \int_{w_2}^{i\infty}  \frac{e^{\frac{\pi
      i}{2}(2k_3-k_4)^2w_1+\frac{3\pi i}{2}k_4^2w_2}}{\sqrt{-(w_1+\tau)^3(w_2+\tau)^3}}dw_1dw_2.
\end{split}
\ee
As a result, we find that the first line of Equation (\ref{R205}) can be expressed as:
\be
\begin{split}
\label{R207}
&-4\pi \sqrt{\frac{3}{y}}\sum_{k_3,k_4}
(2k_3-k_4)M_1(\sqrt{y}(2k_3-k_4)) q^{-\frac{(2k_3-k_4)^2}{4}}\bar
q^{\frac{3}{4}k_4^2}\\
&-\frac{2\sqrt{3}}{y}\bar b_{3,j_\mu}-2\sqrt{3} \sum_{k_3,k_4} \int_{-\bar \tau}^{i\infty} \int_{w_2}^{i\infty}  \frac{e^{\frac{\pi
      i}{2}(2k_3-k_4)^2w_1+\frac{3\pi i}{2}k_4^2w_2}}{\sqrt{-(w_1+\tau)^3(w_2+\tau)^3}}dw_1dw_2,
\end{split}
\ee
where we substituted Equation (\ref{M1period}) for the first integral of
Equation (\ref{R206}).

After addition of the remaining two lines of Equation (\ref{R205}), we
see that the two terms with $\bar b_{3,j_\beta}$ cancel. Moreover, using
the transformation $k_3\mapsto k_4-k_3$, $k_4\mapsto
k_4$, one may show that also the terms with $M_1(\dots)$ cancel. We thus finally arrive at
\be
\label{R208pr}
R^{(2)}_{2,\beta}=-2\sqrt{3}\int_{-\bar \tau}^{i\infty} \int_{w_2}^{i\infty}
\frac{\sum_{k_3,k_4\in \IZ^2+\beta} e^{\frac{\pi
      i}{2}(2k_3-k_4)^2w_1+\frac{3\pi i}{2}k_4^2w_2}}{\sqrt{-(w_1+\tau)^3(w_2+\tau)^3}}dw_1dw_2.
\ee
where we brought the
sum inside the integrand.

\section{Derivation of the completed VW function}
\label{VWP2}
We return in this section to the VW partition functions
$h_{N,\mu}$. After discussing the general structure of the partition
function and its refinement, we give explicit
expressions for gauge groups $U(2)$ and $U(3)$ with fixed 't Hooft fluxes, and derive the
completions using the previous sections. 

\subsection{Refined and unrefined partition functions}
The holomorphic generating functions $h_{N,\mu}(\tau)$ can be determined using
algebraic-geometric techniques \cite{Manschot:2010nc,  Manschot:2014cca, Yoshioka:1994,
  Yoshioka:1995, Gottsche:1990, Manschot:2011ym}. In fact, one arrives
using motivic techniques naturally at a refinement $h_{N,\mu}(\tau,z)$ of
$h_{N,\mu}(\tau)$ (\ref{hN}), where the coefficients of $h_{N,\mu}(\tau)$ are replaced by rational functions of an
additional parameter $w=e^{2\pi i z}$. This extra parameter arises in the partition function by including an
extra fugacity $z$ for the R-symmetry quantum number. The refinement
$h_{N,\mu}(\tau,z)$ is mathematically a
generating function of (weighted) Poincar\'e polynomials rather than
the Euler numbers of the instanton moduli spaces. We will concentrate on the
modular properties of $h_{N,\mu}(\tau)$, since they demonstrate the
 the action of the $SL(2,\IZ)$ $S$-duality group most distinctly.

The function $h_{N,\mu}(\tau)$ is actually only a generating function of
Euler numbers, when the pair $(N,\mu)$ is relatively
prime since the corresponding moduli space of HYM connections (or
semi-stable sheaves) is then compact and smooth. In these cases, there is no ambiguity
in the mathematical interpretation of the VW partition function as a
generating function of Euler numbers or Poincar\'e polynomials of
moduli spaces. However, when the pair $(N,\mu)$
is not coprime, the moduli spaces contain singularities, due to
strictly semi-stable bundles. The topological content of the partition
function is therefore more elusive. It is
conceivable that evaluation of the VW path integral will lead to one of
the various ``types'' of Euler numbers available for such
spaces. While this is hard to carry out directly, we can alternatively avail of
$S$-duality which relates the $h_{N,\mu}$ with different
$\mu$. The $S$-duality relation (\ref{htrafos}) does indeed pick out a
specific rational invariant of the moduli space. Namely, the multiple
cover invariant $\bar \chi(\gamma)$. This {\it rational} invariant is explicitly given as:
\be
\label{barchi}
\bar \chi(\gamma)=(-1)^{\dim_\mathbb{C}\cM_{\gamma}}\sum_{m\geq 1,\, m|\gamma}
(-1)^{\dim_\mathbb{C}\cM_{\gamma/m}}\,\frac{\chi(\gamma/m)}{m^2},
\ee
where the sum runs over positive integers $m\in \mathbb{Z}_{>0}$ which
divide the vector $\gamma$. Note that $(-1)^{\dim_\mathbb{C}\cM_{\gamma}}$ is independent of the
instanton number $n$ and evaluates to
$(-1)^{(N-1)(\mu^2-1)}$ for $\IP^2$. For a generic 't Hooft flux $\mu$, the
exponentiated classical action evaluates to
$q^{n+\frac{\mu^2}{2N}-\frac{N}{8}}$. The VW partition function for gauge group $U(N)$ and
with fixed 't Hooft flux $\mu$, then reads
\be
\label{hNmu}
h_{N,\mu}(\tau)=\sum_{n\in \mathbb{Z}+\frac{\mu}{2}} \bar \chi(\gamma)\,q^{n+\frac{\mu^2}{2N}-\frac{N}{8}}. 
\ee 

The VW partition function $h_{N,\mu}(\tau)$ can be derived from the refined
partition function $h_{N,\mu}(\tau,z)$. The
refined partition function is a generating function of the following
rational functions of $w$:
\be
\Omega(\gamma,w)=\frac{\sum_{\ell=-d}^d b_\ell\, w^{\ell}}{w-w^{-1}},
\qquad w \neq \pm 1, 
\ee
where the positive integer $d$ equals
$\dim_\mathbb{C}\cM_{\gamma}$. The numerator of $\Omega(\gamma,w)$ is a palindromic Laurent
polynomial. More precisely, it is the intersection Poincar\'e polynomial of
the (possibly singular) moduli space of semi-stable sheaves
$\mathcal{M}_\gamma$ multiplied by $w^{-d}$
\cite{Manschot:2016gsx}. The corresponding rational invariant $\bar
\Omega(\gamma,w)$ is defined as:
\be
\bar \Omega(\gamma,w)=\sum_{m|\gamma}\frac{\Omega(\gamma/m,-(-w)^m)}{m}.
\ee
These are the coefficients of the refined partition function $h_{N,\mu}(\tau,z)$:
\be
h_{N,\mu}(\tau,z)=\sum_{n\in \mathbb{Z}+\frac{\mu}{2}} \bar \Omega(\gamma,w)\,q^{n+\frac{\mu^2}{2N}-\frac{N}{8}}.
\ee

One can show for the projective space $\IP^2$, that $\bar \Omega(\gamma,-w)=\pm \bar
\Omega(\gamma,w)$, since all cohomology is even. Rather then taking
the limit $w\to -1$, the numerical invariants $\chi$ and
$\bar \chi$ can therefore be obtained from $\Omega$ and $\bar \Omega$ by taking
the simpler limit $w\to 1$,  
\be 
\label{refinedtonum}
\begin{split}
\chi(\gamma)&=\lim_{w\to 1} (w-w^{-1})\, \Omega(\gamma,w),\\
\bar \chi(\gamma)&=\lim_{w\to 1} (w-w^{-1})\, \bar \Omega(\gamma,w).
\end{split}
\ee

Let us now turn to explicit expressions of the VW partition
functions. For gauge group $U(1)$, the refined partition function $h_{1}(\tau,z)$ is simply the inverse of a
Jacobi theta series \eqref{Jacobitheta} \cite{Vafa:1994tf},
\be
h_{1,0}(\tau,z)=\frac{i}{\theta_1(\tau,2z)},
\ee
In the mathematical literature, $h_{1,0}$ is known as G\"ottsche's formula
\cite{Gottsche:1990} for the cohomology of the Hilbert schemes of
points on $\mathbb{P}^2$. The VW partition function follows straightforwardly,
\be
h_{1,0}(\tau)=\lim_{z\to 0}  4\pi i z \,h_{1,0}(\tau,z)=\frac{1}{\eta(\tau)^3}. 
\ee

Before presenting the explicit expression for $N=2$ and $3$, we
explain the structure for arbitrary $N$. The refined (respectively numerical) VW
partition functions $h_{N,\mu}$ factorize in terms of the $N$'th power
of the $U(1)$ partition function times another function, which we
denote by $g_{N,\mu}(\tau,z)$ (respectively $f_{N,\mu}(\tau)$). We
have more explicitly, 
\be
\begin{split}
&h_{N,\mu}(\tau,z)=g_{N,\mu}(\tau,z)\,h_{1}(\tau,z)^N,\\
&h_{N,\mu}(\tau)=\frac{f_{N,\mu}(\tau)}{\eta(\tau)^{3N}}. 
\end{split}
\ee
One may think of the $g_{N,\mu}$ and $f_{N,\mu}$ in these expressions
as being due to smooth instantons, and $h_{1,0}^N$ as
due the cusps of the moduli space where instantons become
point-like. Note that since
$\lim_{z\to 0} z\, h_{N,\mu}(\tau,z)$ is finite following Equation (\ref{refinedtonum}), and $h_{1,0}(\tau, z)$ has a simple pole at $z=0$,
$g_{N,\mu}(\tau,z)$ has a zero of multiplicity $N-1$ at $z=0$. As a result,
we can write $f_{N,\mu}$ as the $(N-1)$'th derivative of the refined partition function:
\be
\label{ftau}
f_{N,\mu}(\tau)=\frac{1}{(N-1)!}\left(\tfrac{1}{4\pi i }\partial_z \right)^{N-1}
g_{N,\mu}(\tau,z)\vert_{z=0}.
\ee 

The transformation properties of $\eta$ are given in Equation
(\ref{Dedekind}), and we are therefore left with determining the modular
properties of $f_{N,\mu}$ to verify the modularity of the VW partition
function $h_{N,\mu}$. We derive easily from Equation (\ref{htrafos}), that the
{\it expected} transformation properties for the $f_{N,\mu}$ are:
\be
\label{expprops}
\begin{split} 
& f_{N,\mu}\!\left(-\frac{1}{\tau} \right)=\frac{1}{\sqrt{N}}
(-i\tau)^{\frac{3}{2}(N-1)} (-1)^{N-1}\sum_{\nu \mod N} e^{-2\pi i
  \frac{\mu\nu}{N}} f_{N,\nu}(\tau),\\
& f_{N,\mu}(\tau+1)=(-1)^\mu e^{2\pi i\frac{\mu^2}{2N}}
f_{N,\mu}(\tau)
\end{split}
\ee 
Of course, it was established for $N=2$ in Reference \cite{Vafa:1994tf}, that one needs to replace
$f_{2,\mu}$ by a suitable completion $\widehat f_{2,\mu}$ to arrive at
functions which transform as a modular form. We rederive the
completion in the next subsection. A completion is similarly required
for $N=3$, which will be derived in Subsection \ref{secU3}.

\subsection{Gauge group $U(2)$} 
We review in this subsection the VW partition function for $N=2$ and
its modular completion. The refined partition functions are determined
by Yoshioka \cite{Yoshioka:1994, Yoshioka:1995} and equal
\be
\label{U2genfunction}
\begin{split}
g_{2,0}(\tau,z)&=\half +\frac{q^{-\frac{3}{4}}w^5}{\theta_2(2\tau,2z)}\sum_{n\in\Zint}\frac{q^{n^2+n}w^{-2n}}{1-w^4q^{2n-1}}, \\
g_{2,1}(\tau,z)&=\frac{q^{-\frac{1}{4}}w^3}{\theta_3(2\tau, 2z)}\sum_{n\in\Zint}\frac{q^{n^2}w^{-2n}}{1-w^4q^{2n-1}}.
\end{split}
\ee
For later reference, we note that these relations can also be expressed in
terms of $\mu$ and $\Phi$ as
\be
\begin{split}
\label{f2jmu}
g_{2,0}(\tau,z)&=\textstyle \half
  -q^{-\frac{1}{4}}w^{3}\mu(2\tau, 4z-\tau,-2z +\textstyle\frac{1}{2})\\
&=\Phi(\tau,4z,-2z)+\frac{i\,\eta(\tau)^3\theta_3(6\tau,
  6z)}{\theta_1(\tau, 4z)\,b_{3,0}(\tau, 2z)},\\
g_{2,1}(\tau,z) &=-\mu(2\tau, 4z-\tau,-2z-\tau+\textstyle\frac{1}{2})\\
&=w^{-3}q^{-\frac{1}{4}}\left(\Phi(\tau,
  4z,-2z-\tau)-\frac{1}{2}\right) +
\frac{i\,\eta(\tau)^3\theta_2(6\tau, 6z)}{\theta_1(\tau, 4z)\,b_{3,0}(\tau,2z)}.
\end{split}
\ee
Using Equation (\ref{ftau}), one may show that the VW partition functions in
this case are generating functions of the Hurwitz class numbers $H(n)$ \cite{Vafa:1994tf, Bringmann:2010sd}:
\be  
\label{f2j} 
f_{2,\mu}(\tau)=3\sum_{n\geq 0} H(4n-\mu)\,q^{n-\frac{\mu}{4}},\qquad \mu=0,1.
\ee

The modular completions of the refined partition functions $g_{2,\mu}(\tau,z)$ follow from the modular
completion of the $\mu$-function in Equations (\ref{muuv}) and
(\ref{Rutau}). They are given by\footnote{We have chosen to omit
  $\bar \tau$ and $\bar z$ from the arguments of $\widehat g_{2,\mu}(\tau,z)$.}
\be
\begin{split}
&\widehat g_{2,0}(\tau,z)=g_{2,0}(\tau,z) + \half \sum_{\ell \in
  \mathbb{Z}} \left(\sgn(\ell)-E_1(\sqrt{y}(2\ell-3v))\right)q^{-\ell^2}w^{3\ell},\\
&\widehat g_{2,1}(\tau,z)=g_{2,1}(\tau,z) + \half \sum_{\ell \in
  \mathbb{Z}+\half} \left(\sgn(\ell)-E_1(\sqrt{y}(2\ell-3v))\right)\,q^{-\ell^2}w^{3\ell}.
\end{split}
\ee
Note that the $\half$ in the expression for $g_{2,0}(\tau,z)$
(\ref{U2genfunction}) is in fact a non-subleading, holomorphic part of
the completion of $\mu(2\tau,4z-\tau,-2z +\textstyle\frac{1}{2})$, and
should therefore not be considered as part of the subleading,
non-holomorphic terms. Taking the derivative to $z$ at $z=0$, we arrive at the modular completions of $f_{2,\mu}(\tau)$ 
\be
\label{f2complete}
\begin{split}
\widehat
f_{2,\mu}(\tau,\bar \tau)=&f_{2,\mu}(\tau)+\frac{3(1+i)}{8\pi}\int_{-\bar\tau}^{i\infty} \frac{\Theta_{\frac{\mu}{2}}(v)}{(\tau+v)^{\frac{3}{2}}}dv,\\
&=f_{2,\mu}(\tau)-\frac{3i}{4\sqrt{2}\pi}\int_{-\bar\tau}^{i\infty} \frac{\Theta_{\frac{\mu}{2}}(v)}{(-i(v+\tau))^{\frac{3}{2}}}dv,\\
\end{split}
\ee
The $\widehat f_{2,\mu}$ have precisely the expected transformation
properties (\ref{expprops}). As discussed in Section \ref{UNholan},
the non-holomorphic period integral implies an elegant equation for
the holomorphic anomaly (\ref{D2Z2}).

\subsection{Gauge group $U(3)$}
\label{secU3}
Next we move on to $N=3$. Also for this gauge group, there are only
two independent 't Hooft fluxes and therefore only two independent
partition functions, $f_{3,\mu}$ with
$\mu=0,1$. The explicit expressions for the refined partition functions are \cite{Manschot:2014cca}:
\be
\label{g30-1} 
\begin{split}
g_{3,0}(\tau,z)=&\frac{1}{b_{3,0}(\tau,2z)}\sum_{k_1,k_2\in \IZ}
\frac{w^{-2k_1-4k_2}q^{k_1^2+k_2^2+k_1k_2}}{(1-w^4q^{2k_1+k_2})(1-w^4q^{k_2-k_1})}\\
&+\frac{2i
  \eta(\tau)^3}{\theta_1(\tau,4z)\,b_{3,0}(\tau,2z)}\sum_{k\in
  \mathbb{Z}}\frac{w^{-6k}q^{3k^2}}{1-w^6q^{3k}}\\
&-\frac{\eta(\tau)^6\,\theta_1(\tau,2z)}{\theta_1(\tau,4z)^2\,\theta_1(\tau,6z)\,b_{3,0}(\tau,2z)}-g_{2,0}(\tau,z)-\frac{1}{6},
\end{split}
\ee
and
\be
\label{g31-1}
\begin{split}
g_{3,1}(\tau,z)=&\frac{1}{b_{3,0}(\tau,2z)}\sum_{k_1,k_2\in \IZ}
\frac{w^{-2k_1-4k_2+6}q^{k_1^2+k_2^2+k_1k_2-\frac{1}{3}}}{(1-w^4q^{2k_1+k_2-1})(1-w^4q^{k_2-k_1})}\\
&+\frac{i
  \eta(\tau)^3}{\theta_1(\tau,4z)\,b_{3,0}(\tau,2z)}\left(\sum_{k\in
  \mathbb{Z}}\frac{w^{-6k+6}q^{3k^2-\frac{1}{3}}}{1-w^6q^{3k-1}}+\sum_{k\in
  \mathbb{Z}}\frac{w^{-6k}q^{3k^2+3k+\frac{2}{3}}}{1-w^6q^{3k+1}}
\right).
\end{split}
\ee
Note that a few different expressions are available for the
$f_{3,\mu}$, which are related by the blow-up formula
\cite{Manschot:2014cca, BMR2015}. 

The VW partition functions are obtained from Equations (\ref{g30-1})
and (\ref{g31-1}) using Equation (\ref{ftau}).\footnote{Expressions for $f_{3,1}(\tau)$ were also determined using
localization with respect to the toric symmetry of $\IP^2$ by Weist
\cite{WEIST20112406} and Kool \cite{Kool2015}.} The first few coefficients are:
\be
\begin{split}
f_{3,0}(\tau)&=\frac{1}{9}-q+3\,q^2+17\,q^3+41\,q^4+78\,q^5+120\,q^6+O(q^{7}),\\
f_{3,1}(\tau)&=3\,q^{\frac{5}{3}}(1+5\,q+12\,q^2+23\,q^3+38\,q^4+55\,q^5+O(q^6)).
\end{split}
\ee 
See Appendix \ref{Coeff_f3} for explicit expressions for the $q$-series
of these functions, and their first 30 coefficients. These coefficients indicate that their growth is
polynomial. Also note that all coefficients of $f_{3,0}$ are integers,
except the constant term. This is consistent with Equation
(\ref{barchi}), since the coefficients of
$\eta(\tau)^9/(9\,\eta(3\tau)^3)$ are integers, except for the constant
term. In the following we will determine the modular completions of
$f_{3,\mu}$ in a similar way as was discussed for $U(2)$. The modular
properties of the holomorphic $q$-series $f_{3,\mu}$ (\ref{f30shift}) follow
easily from the modular properties of the non-holomorphic terms.
\\

\noi {\bf Completion of $f_{3,0}$}\\ 
The completion of the refined partition function $g_{3,0}(\tau,z)$ is given by:
\be
\label{r3nonholo}
\widehat
g_{3,0}(\tau,z)=g_{3,0}(\tau,z)-\frac{1}{12}-\frac{1}{4}R_{2,0}(\tau,6z)-\sum_{\mu=0,1} g_{2,\mu}(\tau,z)\,R_{1,\frac{\mu}{2}}(\tau,6z).
\ee
while the completion of the VW partition function $f_{3,0}(\tau)$ is given:
\be
\label{f30complete}
\widehat f_{3,0}(\tau,\bar \tau)=f_{3,0}(\tau)-\frac{i}{\pi}\left(\frac{3}{2} \right)^{\frac{3}{2}}\sum_{\mu=0,1}\int_{-\bar \tau}^{i\infty}
\frac{\widehat
  f_{2,\mu}(\tau,-v)\,\Theta_{\frac{\mu}{2}}(3v)}{(-i(v+\tau))^{\frac{3}{2}}}dv.
\ee

\noi {\bf Proof}\\
We start by bringing $g_{3,0}(\tau,z)$ in a more convenient form by
substituting Equation (\ref{f2jmu}) for $g_{2,0}(\tau,z)$ in Equation (\ref{g30-1}). After combining
the first term on the rhs of $g_{2,0}(\tau,z)$ in Equation (\ref{f2jmu}) with
the first term on the rhs of Equation (\ref{g30-1}), one can express $g_{3,0}$ as
\be
\label{f30} 
\begin{split}
g_{3,0}(\tau,z)=&\,\frac{1}{4}+\frac{w^4}{b_{3,0}(\tau, 2z)}
\sum_{k_1,k_2\in \IZ}
\frac{w^{-2k_1-4k_2}q^{k_1^2+k_2^2+k_1k_2+2k_1+k_2}}{(1-w^4q^{2k_1+k_2})
  (1-w^4q^{k_2-k_1})} \\
&+\frac{2i\,\eta(\tau)^3}{\theta_1(\tau,4z)\,b_{3,0}(\tau,2z)}\left(-\tfrac{1}{2}\theta_3(6\tau,
  6z)+ \sum_{k\in
  \mathbb{Z}}\frac{w^{-6k}q^{3k^2}}{1-w^6q^{3k}}\right) \\
&-\frac{\eta(\tau)^6\,\theta_1(\tau,2z)}{\theta_1(\tau,4z)^2\,\theta_1(\tau,6z)\,b_{3,0}(\tau,2z)}+\frac{1}{12}.
\end{split}
\ee    
 
We can now determine
the completion $\widehat f_{3,0}$ by determining the completions using the
results of Section \ref{AppellLerch}. Working line by line, we arrive
at the following completions: 
\begin{enumerate} 
\item The first line of the rhs of Equation (\ref{f30}) equals $\Psi_0(4z,-2z)$, a
  specialization of $\Psi_0(u,v)$ which is defined in
  (\ref{Psi01}). The completion of $\Psi_0(u,v)$ is given by Equation
  (\ref{RPsi0}). Specialization of the latter, provides the completion of the first line:
\be
-\frac{1}{4}R_{2,0}(6z)-\Phi(4z,-2z)\,R_{1,0}(6z)-w^{-3}q^{-\frac{1}{4}}
\left(\Phi(4z+\tau,-2z)-\frac{1}{2}\right)\,R_{1,1}(6z).
\ee 
\item The completion of the second line follows from Equation (\ref{Rutau}), and equals
\be
-\frac{i\eta(\tau)^3}{\theta_1(\tau, 4z)\,b_{3,0}(\tau, 2z)}\left(\theta_3(6\tau, 6z)R_{1,0}(6z)+\theta_2(6\tau, 6z)R_{1,1}(6z) \right)
\ee
\item The first term of the third line transforms as a Jacobi form of
  weight 1 and index $-36$ and does therefore not require a
  completion. However, the constant term $\frac{1}{12}$ does not
  transform appropriately. Therefore we subtract it, such that the
  completion of the third line is
\be
-\frac{1}{12}.
\ee
\end{enumerate}
Adding the three contributions above and substitution of the $N=2$ partition
functions $g_{2,\mu}$ using Equation (\ref{f2jmu}), we find the
claimed expression in Equation (\ref{r3nonholo}).

Our next aim is to determine the modular completion of the VW
partition function. To this end, we make a Taylor expansion of
$\widehat g_{3,0}(\tau,z)$ around $z=0$. We already discussed that the
constant and linear term of $g_{3,0}(\tau,z)$ vanishes. This is in fact also
the case for $\widehat g_{3,0}$. To see this, note that the constant
$-\frac{1}{12}$ in Equation (\ref{r3nonholo}) cancels agains the constant term of $R_{2,\mu}$
(\ref{R208}). Furthermore, the $g_{2,\mu}$ and $R_{1,\alpha}$ in
Equation (\ref{r3nonholo}) all start with a linear term in $z$. 

Let us now determine the completion $\widehat f_{3,0}(\tau)$ by
determing the quadratic term in the Tayler expansion of $\widehat g_{3,0}(\tau,z)$ using the results of
Section \ref{Taylerexp}. Substitution of Equations (\ref{R1j}) and (\ref{R208}) for
$R^{(1)}_{1,\alpha}$ and $R_{2,0}^{(2)}$ in the $\frac{1}{2}\partial^2_z
\widehat g_{3,0}(\tau,z)|_{z=0}$, gives 
\be
\begin{split}
\widehat f_{3,0}(\tau,\bar \tau)=&f_{3,0}(\tau) +\frac{6}{4\pi i}\sum_{\mu=0,1}
f_{2,\mu}(\tau) \sqrt{\frac{3}{2}} \sum_{\ell\in
  \mathbb{Z}+\frac{\mu}{2}} \int_{-\bar \tau}^{i\infty} \frac{e^{6\pi i
    \ell^2 w}}{(-i(w+\tau))^{\frac{3}{2}}}dw\\
&+\frac{36}{(4\pi i)^2}\frac{\sqrt{3}}{2}\int_{-\bar \tau}^{i\infty} \int_{w_2}^{i\infty}
\frac{\sum_{k_3,k_4\in \IZ} e^{\frac{\pi
      i}{2}(2k_3-k_4)^2w_1+\frac{3\pi
      i}{2}k_4^2w_2}}{\sqrt{-(w_1+\tau)^3(w_2+\tau)^3}}dw_1dw_2.
\end{split} 
\ee
We split the sum over $k_3$ and $k_4$ on the second line into a sum
with $k_4$ even and one with $k_4$ odd. Then the sum splits into a
sum of products of theta series:
\be
\sum_{k_3,k_4\in \IZ} e^{\frac{\pi 
      i}{2}(2k_3-k_4)^2w_1+\frac{3\pi
      i}{2}k_4^2w_2}=\sum_{\mu=0,1}\Theta_{\frac{\mu}{2}}(w_1)\, \Theta_{\frac{\mu}{2}}(3w_2),
\ee
with $\Theta_\alpha$ as in Equation (\ref{Theta_alpha}). Then using
Equation (\ref{f2complete}), we can further simplify $\widehat
f_{3,0}$ to the expression in Equation (\ref{f30complete}). \hfill $\square$
\\

\noi {\bf Completion of $f_{3,1}$}\\
The completion of the refined partition function $g_{3,1}(\tau,z)$ is given by:
\be
\label{hatf31}
\begin{split}
\widehat
g_{3,1}(\tau,z)=&g_{3,1}(\tau,z)-\frac{1}{4}R_{2,\frac{1}{3}(-1,1)}(6z)\\
&  -\frac{1}{2}\, g_{2,0}(\tau,z)\,(R_{1,\frac{1}{3}}(6z)-R_{1,\frac{1}{3}}(-6z))    \\
&  -\frac{1}{2}\, g_{2,1}(\tau,z)\,(R_{1,\frac{1}{6}}(6z)-R_{1,\frac{1}{6}}(-6z)).
\end{split}
\ee
From a Taylor expansion of $\widehat g_{3,1}$, one can derive that the
completion $\widehat f_{3,1}$ of the VW partition function $f_{3,1}$
is given by:
\be
\label{f31complete}
\widehat f_{3,1}(\tau,\bar \tau)=f_{3,1}(\tau)-\frac{i}{\pi}\left(\frac{3}{2} \right)^{\frac{3}{2}}\sum_{\mu=0,1}\int_{-\bar \tau}^{i\infty}
\frac{\widehat
  f_{2,\mu}(\tau,-v)\,\Theta_{\frac{1}{3}+\frac{\mu}{2}}(3v)}{(-i(v+\tau))^{\frac{3}{2}}}dv. 
\ee

\noi {\bf Proof}\\
The derivation of the completion of $f_{3,1}(\tau)$ is similar to the one
of $f_{3,0}(z)$ discussed above. The explicit expression for the
refined generating function $f_{3,1}(\tau,z)$ was given in Equation
(\ref{g31-1}). Section \ref{AppellLerch} provides the completion of
each line as before:
\begin{enumerate}
\item The first line of Equation (\ref{g31-1}) equals the Appell-Lerch
  $\Psi_1(4z,-2z)$ with signature $(2,2)$ defined in Equation (\ref{Psi01}). The additional non-holomorphic terms follow by
  specializing $R_{\Psi_1}$ (\ref{RPsi1}):
\be
\begin{split}
&-\frac{1}{4}R_{2,\frac{1}{3}(-1,1)}(6z)-\frac{1}{2}\Phi(4z,-2z)(R_{1,\frac{1}{3}}(6z)-R_{1,\frac{1}{3}}(-6z))\\
& -\frac{1}{2} w^{-3}q^{-\frac{1}{4}}\left(
  \Phi(4z+\tau,-2z)-\tfrac{1}{2}   \right)\left( R_{1,\frac{1}{6}}(6z)-R_{1,\frac{1}{6}}(-6z)\right).
\end{split} 
\ee
\item The second line of Equation (\ref{g31-1}) can be expressed in
  terms of $A_{1}$ and $A_2$ defined in Equation (\ref{Adef}). The
  additional terms for the completion can be read off from Equation
  (\ref{Aicomp}) and read: 
\be
\begin{split}
&-\frac{1}{2} \frac{i\eta^3}{\theta_1(\tau,4z)\,b_{3,0}(\tau,2z)}\\
&\times \left(\theta_3(6\tau,6z)
  (R_{1,\frac{1}{3}}(6z)-R_{1,\frac{1}{3}}(-6z)) +  \theta_3(6\tau,6z) (R_{1,\frac{1}{6}}(6z)-R_{1,\frac{1}{6}}(-6z))\right).
\end{split}
\ee
\end{enumerate}
Adding these terms to the holomorphic part $g_{3,1}$, and
substitution of Equation (\ref{f2jmu}) gives for $\widehat g_{3,1}(\tau,z)$
the expression in Equation (\ref{hatf31}). The completion of the VW partition function $f_{3,1}(\tau)$ is obtained as
for $f_{3,0}(\tau)$ above, by determining $\frac{1}{2}\partial^2_z
\widehat f_{3,1}(\tau,z)|_{z=0}$. One arrives in this way at the expression in
Equation (\ref{f31complete}). \hfill $\square$

\subsection*{Acknowledgements} 
I would like to thank Sergey Alexandrov, Sibasish Banerjee and Boris
Pioline for collaborating on related subjects and in particular on Reference
\cite{Alexandrov:2016enp}. I moreover wish to thank Sergey Alexandrov,
Kathrin Bringmann, Greg Moore, Boris Pioline, Cumrun Vafa and Sander Zwegers for 
discussions and correspondence. 
\newpage 
\appendix  

\section{Explicit $q$-series for $f_{3,\mu}$}
\label{Coeff_f3}

\subsection{$q$-series}
The $q$-series $f_{3,\mu}$ is defined in terms of the
$g_{3,\mu}(\tau,z)$ by Equation (\ref{ftau}). Based on the explicit expressions for $g_{3,\mu}(\tau,z)$,
(\ref{g31-1}) and (\ref{f30}), we can derive
explicit $q$-series for $f_{3,\mu}(\tau)$. To this end, recall the
classical Eisenstein series $E_k(\tau)$ of weight $k\in 2\mathbb{N}$, which have the
$q$-expansion
\be
E_k(\tau)=1-\frac{2k}{B_k}\sum_{n=1}^\infty
\frac{n^{k-1}q^n}{1-q^n},
\ee
where $q=e^{2\pi i \tau}$, and $B_k$ are the Bernoulli numbers, $B_2=\frac{1}{6}$,
$B_4=-\frac{1}{30},$ etc. We define furthermore the
following summands
\be
\label{S2S11}
\begin{split}
S_{1,\mu}(k;q)=&\,\frac{(-1+E_2(\tau))(k-\mu+1)}{2(1-q^{3k-\mu})}+\frac{9(k-\mu)^2+33(k-\mu)+31-E_2(\tau)}{2(1-q^{3k-\mu})^2}\\
&-\frac{15(k-\mu)+34}{(1-q^{3k-\mu})^3}+\frac{19}{(1-q^{3k-\mu})^4},\\
S_{2}(A,B;q)=&\,\frac{4\, q^B}{(1-q^A)(1-q^B)^3}+\frac{4\,q^A}{(1-q^A)^3
  (1-q^B)} + \frac{4}{(1-q^A)^2 (1-q^B)^2} \\
&-\frac{2(A+B+1)q^B}{(1-q^A)
    (1-q^B)^2} -\frac{2(A+B+1)q^A}{(1-q^A)^2(1-q^B)} +\frac{(A+B-2)^2-8}{2(1-q^A)(1-q^B)} . 
\end{split}
\ee
We then have:
\begin{itemize}
\item 
$b_{3,0}\,f_{3,0}$ can be expressed as:
\be
\label{qseriesf30}
\begin{split}
b_{3,0}(\tau)\,f_{3,0}(\tau)=&\,\frac{13}{240}+\frac{1}{24}E_2(\tau)+\frac{1}{72}E_2(\tau)^2+\frac{1}{720}E_4(\tau) \\
&-\frac{9}{2} \sum_{k\in \IZ} k^2
q^{3k^2}+\frac{1}{6}\sum_{k_1,k_2\in \IZ} (k_1+2k_2)^2q^{k_1^2+k_2^2+k_1k_2} \\
&+\sum_{k\in \IZ \atop k\neq 0} S_{1,0}(k;q)\,q^{3k^2}\\
&+\sum_{k_1,k_2\in \IZ\atop {
2k_1+k_2\neq 0,\atop k_2\neq k_1}}S_{2}(2k_1+k_2,k_2-k_1;q)\,q^{k_1^2+k_2^2+k_1k_2+2k_1+k_2},
\end{split}
\ee
\item $b_{3,0}\,f_{3,1}$ can be expressed as:
\be
\label{qseriesf31}
\begin{split}
b_{3,0}(\tau)\,f_{3,1}(\tau)=&\sum_{k\in \IZ } S_{1,1}(k;q)\,q^{3k^2-\frac{1}{3}}\\
&+\sum_{k_1,k_2\in \IZ\atop
2k_1+k_2\neq 1, k_2\neq k_1}S_{2}(2k_1+k_2-1,k_2-k_1;q)\,q^{k_1^2+k_2^2+k_1k_2-\frac{1}{3}}.
\end{split}
\ee
\end{itemize}

\noi {\bf Proof}\\
We prove Equation (\ref{qseriesf31}) by determining the coefficient of
$z^2$ in the Taylor expansion of
$b_{3,0}(\tau,z)\,g_{3,1}(\tau,z)$. We rewrite Equation
(\ref{g31-1}) as 
\be
\label{b30g1tauz}
\begin{split}
b_{3,0}(\tau,z)\,g_{3,1}(\tau,z)=&\sum_{k_1,k_2\in \IZ\atop 2k_1+k_2\neq 1; k_2\neq k_1}
\frac{w^{-2k_1-4k_2+6}q^{k_1^2+k_2^2+k_1k_2-\frac{1}{3}}}{(1-w^4q^{2k_1+k_2-1})(1-w^4q^{k_2-k_1})}\\
&+\frac{1}{1-w^4}\left(
  \frac{w^{-6k+6}q^{3k^2-\frac{1}{3}}}{1-w^4q^{3k-1}}+
  \frac{w^{-6k+2}q^{3k^2+3k+\frac{2}{3}}}{1-w^4q^{3k+1}} \right)\\
&+\frac{i\eta(\tau)^3}{\theta_1(\tau,z)}\left(\sum_{k\in
  \IZ} \frac{w^{-6k+6}q^{3k^2-\frac{1}{3}}}{1-w^6q^{3k-1}} + \sum_{k\in
  \IZ} \frac{w^{-6k}q^{3k^2+3k+\frac{2}{3}}}{1-w^6q^{3k+1}}\right),
\end{split}
\ee
where the first line is finite in the limit $z\to 0$. The second
and third line have a first order pole, which cancel each other. The Taylor coefficient of
$z^2$ of $\frac{e^{-(2A+2B-4)z}}{(1-e^{4z}q^{A})(1-e^{4z}q^B)}$ with
$A,B\neq 0$ is given by $4\,S_2(A,B;q)$ (\ref{S2S11}), which
reproduces the 
second line of Equation (\ref{qseriesf31}). Using the Laurent expansion
\be
\frac{i\,\eta^3(\tau)}{\theta_1(\tau,z)}=\frac{1}{2\pi i
  z}-\frac{1}{24}E_2(\tau)\,(2\pi i z)+ \frac{1}{5760} (5E_2(\tau)^2+2E_4(\tau)) (2\pi i z)^3+\dots,
\ee
one can verify that the coefficient of $(2\pi i z)^2$
in the Taylor expansion of the sum of the second and third line of
(\ref{b30g1tauz}) is $4\,S_{1,1}(k;q)$. We thus arrive at the claimed result for $f_{3,1}$.

To determine $f_{3,0}$, one groups terms of $b_{3,0}\,g_{3,0}$ which are finite for $z\to
0$, or have poles of order one and two.  The claimed expression for
$f_{3,0}$ (\ref{qseriesf30}) follows then in a
similar way for $f_{3,1}$.  \hfill $\square$
 
\subsection{Coefficients}

Let $d_\mu(n)$ be the coefficients of $f_{3,\mu}$, $\mu=0,1$, defined as
\be
f_{3,\mu}(\tau)=\sum_{n\geq 0} d_{\mu}(n)\,q^{n-\frac{\mu}{3}}.
\ee
Using Equations (\ref{qseriesf30}) and (\ref{qseriesf31}) it is easy to determine the
first $d_{\mu}(n)$. Table \ref{tab:coeffs} lists the first 30 coefficients $d_{\mu}(n)$ 
for $\mu=0$ and 1.
 
\begin{table}[h!]
\centering
\begin{tabular}{l|rr}
\hline
$n$ & $d_{0}(n)$ & $d_1(n)$ \\
\hline
0 & $\tfrac{1}{9}$& 0 \\
1 & -1 & 0\\
2 & 3 & 3 \\
3 & 17 & 15 \\
4 & 41 & 36 \\
5 & 78 & 69 \\
6 & 120 & 114 \\
7 & 193 & 165 \\
8 & 240 & 246 \\
9 & 359 & 303 \\
10 & 414 & 432\\
11 & 579 & 492 \\
12 & 626 & 669 \\
13 & 856 & 726 \\
14 & 906 & 975 \\
15 & 1194 & 999 \\
16 & 1172 & 1332 \\
17 & 1638 & 1338 \\
18 & 1569 & 1743 \\
19 & 1987 & 1716 \\
20 & 2040 & 2226 \\
21 & 2578 & 2130 \\
22 & 2340 & 2775 \\
23 & 3255 & 2625 \\
24 & 2940 & 3354\\
25 & 3665 & 3129 \\
26 & 3642 & 4041 \\
27 & 4490 & 3735 \\
28 & 3940 & 4752 \\
29 & 5484 & 4317 \\
30 & 4734 & 5532\\
\end{tabular}
\caption{First 30 coefficients of the functions $f_{3,\mu}$, $\mu=0,1$, discussed
  in the main text.}
\label{tab:coeffs}
\end{table}

\end{document}